\documentclass[11pt,a4paper]{article}
\pdfoutput=1
\usepackage[english]{babel}
\usepackage{amsfonts,amsbsy,bm,euscript,mathrsfs,bbm}
\usepackage{amssymb,stmaryrd,faktor,slashed}
\usepackage[x11names]{xcolor}
\usepackage[tbtags]{amsmath}
\usepackage[bookmarks=true,colorlinks=true,linkcolor=black,citecolor=black,urlcolor=black,bookmarksnumbered]{hyperref}
\usepackage[nosort]{cite}
\usepackage{tensor,braket}
\usepackage{tablefootnote}
\usepackage{graphicx}
\usepackage[normalem]{ulem}

\numberwithin{equation}{section}

\makeatletter
\renewcommand\section{\@startsection {section}{1}{\z@}
{-3.5ex \@plus -1ex \@minus -.2ex}
{2.3ex \@plus.2ex}
{\normalfont\Large\bfseries}}
\renewcommand\subsection{\@startsection{subsection}{2}{\z@}
{-3.25ex\@plus -1ex \@minus -.2ex}
{1.5ex \@plus.2ex}
{\normalfont\large\bfseries}}
\makeatother

\newcommand{\abs}[1]{\left\vert#1\right\vert}

\newcommand{\tr}{\mathrm{tr}}

\newcommand{\form}[1]{\mathrm{d}#1}
\newcommand{\scom}[2]{\left[#1\stackrel{\star}{,}#2\right]}

\begin{document}

\setcounter{equation}{0}
\setcounter{footnote}{0}
\setcounter{section}{0}

\thispagestyle{empty}

\begin{flushright}
\texttt{}
\end{flushright}

\begin{center}

\vspace{1.5truecm}

{\LARGE \bf Twisting Twistor space}

\vspace{1.5truecm}

{Tim Meier$^{1}$ and Eggon Viana$^{1,2,3}$}

\vspace{1.0truecm}

{\em $^1$ Instituto Galego de Física de Altas Enerxías (IGFAE),\\
and Departamento de Física de Partículas,\\
Universidade de Santiago de Compostela,\\
15705 Santiago de Compostela, Spain}

\vspace{.5cm}

{\em $^2$ ICTP South American Institute for Fundamental Research\\
Instituto de F\'{i}sica Te\'{o}rica, UNESP-Universidade Estadual Paulista\\
R. Dr. Bento T. Ferraz 271, Bl. II, S\~{a}o Paulo 01140-070, SP, Brazil\\
}

\vspace{.5cm}

{\em $^3$ Department of Mathematical Sciences,\\
Durham University, Durham DH1 3LE, UK}

\vspace{1.0truecm}

{{\tt tim.meier@usc.es // eggon.viana@unesp.br}}

\vspace{1.0truecm}
\end{center}

\begin{abstract}
We construct twisted noncommutative gauge theories on twistor space and show that they are equivalent to four-dimensional twist-noncommutative gauge theories. In particular, we study twists of the Poincar\'e algebra. We explain how such a twist leads to twisted noncommutative twistor space and how to construct noncommutative versions of BF theory and holomorphic Chern-Simons theory on noncommutative supertwistor space. We show how those theories are equivalent to noncommutative versions of Yang-Mills theory and supersymmetric Yang-Mills theory, respectively.

\end{abstract}

\newpage

\setcounter{equation}{0}
\setcounter{footnote}{0}
\setcounter{section}{0}

\tableofcontents

\section{Introduction}
Twistor space provides a powerful framework for reformulating four-dimensional field theories in terms of complex geometry on projective space, replacing the notion of a spacetime point with holomorphic structures. This perspective has proved particularly fruitful in the study of scattering amplitudes, where twistor-inspired methods reveal underlying symmetries, such as dual conformal invariance, that remain concealed in conventional Feynman diagram expansions \cite{Amp:Roiban,Amp:Roiban2,Amp:Roiban3,Amp:Cachazo,Amp:Bern,Amp:Gukov}. From this point of view, the twistor-space formulation of gauge theories offers a natural setting for gaining deeper insight into the geometric and symmetry principles underlying scattering amplitudes.

Beyond amplitudes, twistor space offers a natural setting for constructing gauge theories through topological models. Holomorphic bundles over twistor space encode the self-dual Yang-Mills equations, derivable from BF theory \cite{Adamo:Lectures}, while Chern-Simons theory on twistor space underpins the twistor string approach \cite{Witten:CS}. This framework not only connects gauge theory to topological string theory, but also illuminates its links to integrable systems.

Recent progress revealing twistor theory as a natural generation tool of two‑dimensional integrable models provides an additional motivation for exploring gauge theories on twistor space. In particular, recent developments have shown that a wide class of integrable field theories can be obtained from holomorphic Chern–Simons or BF-type gauge theories. These constructions for example reveal a unifying “diamond” of theories interpolating between 6D holomorphic Chern–Simons theory, 4D gauge theories, and 2D integrable models \cite{Ryan:Twistor,Ben:Diamond,Lewis:BFTheory,Thompson:Diamond}. From this perspective, twistor space plays a central role, offering a geometric setting in which higher-dimensional gauge theories and their integrable reductions can be studied in a systematic manner.

A particularly striking application arises in the context of $\mathcal{N}=4$ Super Yang-Mills theory and AdS/CFT. In \cite{Witten}, a string-theory dual to perturbative $\mathcal{N}=4$ SYM theory was considered, given by a topological string theory on the supertwistor space $\mathbb{CP}^{3|4}$, known as the B-model \cite{Witten:CS,Witten:Topological}. More specifically, the anti-self-dual sector of $\mathcal{N}=4$ SYM is captured by a holomorphic Chern-Simons theory on $\mathbb{CP}^{3|4}$, which describes the open string sector of this topological B-model. Perturbations around this sector, implemented via nonlocal interaction terms on twistor space, reconstruct the full spacetime action of $\mathcal{N}=4$ SYM \cite{Mason,Mason:twist,Adamo:Lectures}, up to a topological term.

In this paper, we study twisted noncommutative deformations of the BF theory on $\mathbb{CP}^3$ and its supersymmetric cousin, the holomorphic Chern-Simons theory on the supertwistor space $\mathbb{CP}^{3|4}$ with the aim of deepening the understanding of scattering amplitudes in the context of noncommutative gauge theories as well as its connection to integrable and noncommutative integrable models in the spirit of \cite{Takasaki:2000vs,Hamanaka:2005mq}. Another motivation for this study is the natural
formulation of MHV diagram rules to calculate scattering amplitudes in Yang-Mills theories through the formulation of the corresponding twistor theories \cite{Adamo:MHV}. With our framework we are aiming on providing a framework for similar results in the setup of noncommutative gauge theories.

In this paper we will start by reviewing the construction of BF theory--corresponding to pure anti-self-dual Yang-Mills theory--and supersymmetric holomorphic Chern-Simons theory in Section~\ref{sec:TwistorSpace} before we turn to noncommutative extensions. The idea of noncommutative spacetime dates back to early quantum field theory as a regulator for quantum observables~\cite{Szabo,Madore, Snyder}. It was later argued that spacetime itself may become quantized at the Planck scale, providing a framework for quantum gravity~\cite{Fredenhagen:QG,Arzano:QG,Addazi:QG}. Noncommutative gauge theories also arise as low-energy limits of open strings in the presence of a nonconstant $B$-field~\cite{Seiberg, Gopakumar,Seiberg-Witten, Douglas}, and hence in the context of AdS/CFT holography~\cite{Maldacena,Maldacena-Russo,Hashimoto}.

A consistent way to lift observables from smooth commutative spacetime to operator-valued noncommutative spacetime is given by Weyl quantization~\cite{Szabo,Madore}. The corresponding Wigner-Weyl map provides a framework to change the point of view from operator-valued objects living on spacetime with operator-valued noncommutative coordinates to ordinary fields on ordinary spacetime coordinates, where the noncommutativity is implemented via a noncommutative product, the star product~\cite{Kontsevich}. Although the Groenewold-Moyal product corresponds to constant noncommutativity~\cite{Snyder}, these ideas extend to a broad class of deformations, including nonconstant cases. In this work, we focus on star products generated via Drinfel'd twists, which offer clear algebraic properties and a differential calculus~\cite{Drinfeld}. Such deformations have been studied in various contexts. There is a large class of twisted versions of $\mathcal{N}=4$ SYM which admit twisted superconformal symmetry at least at the classical level \cite{Lunin,Meier:twist,Borsato&Meier}. In the context of the AdS/CFT correspondence, so called homogeneous Yang-Baxter deformations of the $AdS_5 \times S^5$ string have been studied in recent years \cite{Delduc:2013qra,vanTongeren:2015soa,Kawaguchi:2014qwa,Matsumoto:2015jja}. These are conjectured to be dual to twist deformations of the $\mathcal{N}=4$ SYM theory~\cite{vanTongeren:twist, Vicedo:twist, Borsato:twist}. Furthermore, the famous $\kappa$-Minkowski spacetime can be expressed via a Drinfel'd twist built from the vector field corresponding to the dilatation operator and a translation, however the formulation of gauge invariant Yang-Mills actions for a finite deformation parameter is subtle \cite{Dimitrijevic:2014dxa}. More recently, gauge invariant twist-noncommutative gauge theories have been constructed first for twists in the Poincar\'e algebra \cite{Meier:twist,Meier:quad,Meier:SpinChain} and has been generalized to allow the dilatation generator in the twist as well \cite{Borsato&Meier}.

On the other hand, a different class of deformations arises in the context of gauge theories with less supersymmetries. An important example, known as $\beta$ deformation, is obtained as an exactly marginal deformation of $\mathcal{N}=4$ SYM that breaks the $R$-symmetry algebra \cite{Leigh, Fokken}. Although, the deformation is a twist of only the internal R-symmetry of the model, it can be viewed as being generated by a noncommutative product between the individual fields. As such, the $\beta$ deformation fits the general noncommutative field theory construction discussed in \cite{Meier:twist,Borsato&Meier} as well. It has a gravity dual, described by the Maldacena-Lunin background \cite{Lunin, Frolov}. The corresponding string description of this deformation has been studied in \cite{Alday}, and a covariant approach using the pure spinor formalism was developed in \cite{Andrei, Benitez, Viana2}. Following the correspondence between topological string theory on supertwistor space and perturbative $\mathcal{N}=4$ SYM \cite{Witten}, specific deformations of the topological B-model on supertwistor space have been shown to encode deformations of its dual gauge theory \cite{Hofman,Hofman:Bmodel,Viana}. In this direction, a twistor-space formulation of the $\beta$-deformed gauge theories were studied in \cite{Kulaxizi, Gao, Adamo:fish}. By contrast, a twistorial formulation of noncommutative gauge theories has not been developed beyond the case of the simplest Groenewold-Moyal-type noncommutativity discussed in \cite{Takasaki:2000vs,Hamanaka:2005mq}.

Motivated by these developments, we construct a twisted twistor space for Drinfel'd twists in the Poincar\'e algebra and gauge theories thereon. Section~\ref{sec:DefBFTheory} reviews gauge-invariant Yang-Mills theory in four dimensions, while Section~\ref{sec:DefTwistor} presents our main results on twisted twistor space. In particular, Section~\ref{subsec:NCTwistor} introduces the twist deformation of twistor space. This construction incorporates a twisted notion of Penrose transformations related to the one discussed in \cite{Majid&Brain}. In addition, we will discuss a consistent way to define gauge theories thereon and relate those to noncommutative four-dimensional gauge theories. In particular, we construct a deformation of the BF theory, which is a noncommutative gauge theory in Section~\ref{subsec:twistorYM} and its supersymmetric extension in Section~\ref{subsec:twistorSYM}. We also establish the equivalence between the deformed theories on twistor space and their counterparts on four-dimensional spacetime by defining a fibration of the projective space into euclidean noncommutative spacetime and a residual $\mathbb{CP}^1$ space. Our approach is heavily based on leaving structures on the fibre--such as volume forms and basis forms--undeformed and star-commutative. As a result, we restrict our construction to twists in the Poincar\'e algebra. This class of deformations is well understood in terms of constructing gauge invariant spacetime field theories and hence, we will generalize the equivalence between gauge theories on twistor space and four-dimensional spacetime to their twist-deformed counterparts.

\section{Gauge Theories on Twistor Space}\label{sec:TwistorSpace}

In this section, we begin with a review of twistor-space geometry, introducing a convenient parameterization of the space and its natural supersymmetric extension. Within this framework, conformal transformations acquire a particularly transparent description, namely they are linear transformations on twistor space. This elegant realization of the conformal algebra will prove useful in what follows. Building on this structure, we then revisit the twistor action for Yang–Mills theory and its $\mathcal{N}=4$ supersymmetric extension.

\subsection{Twistor space}

The twistor space $\mathbb{PT}^3$ of complexified Minkowski space is defined to be an open subset of the complex projective space $\mathbb{CP}^3$. It is a three-dimensional complex projective space, that can be understood being generated by homogeneous coordinates $Z^I$, for $I=1,\dots,4$. In order to reduce this $\mathbb{C}^4$ to $\mathbb{CP}^3$, points are identified with each other if they are on the same line passing the origin and hence
\begin{equation}
    Z^I\sim tZ^I\qquad \forall t\in \mathbb{C}.
\end{equation}
The scaling behavior of a function or differential form on twistor space under this identification is called homogeneous degree. For example, the coordinates themselves are of homogeneous degree 1. For our purpose of identifying the twistor space with four dimensional spacetime, we will separate the four homogeneous coordinates $Z^i$ into two Weyl spinors of opposite chirality:
\begin{equation}
Z^I = (\lambda_\alpha,\mu^{\dot\alpha}).
\end{equation}
The relation between $\mathbb{CP}^3$ and spacetime is captured by an algebraic relation between
the coordinates $Z^I$ on twistor space and the coordinates $x^{\alpha\dot\alpha}$ on complexified Minkowski space:
\begin{equation}\label{eq:incidence}
\mu^{\dot\alpha}=\lambda_\alpha x^{\dot\alpha\alpha}.
\end{equation}
This relation is referred to as \textit{twistor correspondence}, and the equation above will be refered to as the \textit{incidence relation}.

We now introduce a reality structure on the complexified Minkowski space in order to recover a four-dimensional real spacetime. This requires specifying  a complex conjugation on twistor space. In this work, we use the following conjugation:
\begin{equation}
\mu^{\dot\alpha} = (\mu^{\dot0},\mu^{\dot1}) \mapsto \hat\mu^{\dot\alpha}\equiv(-\bar\mu^{\dot1},\mu^{\dot0}),\ \ \ \ \ \ \ \ \ \lambda_\alpha = (\lambda_0,\lambda_1) \mapsto \hat\lambda_\alpha\equiv(-\bar\lambda_1,\bar\lambda_0).
\end{equation}
Note that by choosing a reality condition, a particular real slice in the complex Minkowski space is chosen. Such a choice corresponds to a particular signature in the resulting real spacetime \cite{PenroseRindlertwistor}. The above choice corresponds to the euclidean signature, as we are interested in the euclidean spacetime. This conjugation induces a $\mathbb{CP}^1$-fibration over the euclidean spacetime: $\mathbb{PT}^3\to\mathbb{R}^4$. This means that every point of twistor space gets mapped to a point in $\mathbb{R}^4$ using the reality condition, while each point of $\mathbb{R}^4$ corresponds to a twistor line $L_x\cong\mathbb{CP}^1$. In coordinates, the twistor line is parametrized by $\lambda_\alpha$: $L_x=\{ (\lambda_\alpha,\mu^{\dot\alpha})\in\mathbb{CP}^3\ ;\ \mu^{\dot\alpha}=x^{\dot\alpha\alpha}\lambda_\alpha\}$, while the fibration is given explicitly by:
\begin{equation}\label{x=lm}
Z^I = (\lambda_\alpha,\mu^{\dot\alpha}) \to x^{\dot\alpha\alpha} = \frac{\hat\mu^{\dot\alpha}\lambda^\alpha - \mu^{\dot\alpha}\hat\lambda^\alpha}{\langle\lambda\hat\lambda\rangle},
\end{equation}
where $\langle\lambda\hat\lambda\rangle = \epsilon^{\alpha\beta}\lambda_\beta\hat\lambda_\alpha$, $\lambda^\alpha = \epsilon^{\alpha\beta}\lambda_\beta$ and $\hat\lambda^\alpha = \epsilon^{\alpha\beta}\hat\lambda_\beta$. Thus, in Euclidean signature, a point in twistor space is specified by fixing two points: one in $\mathbb{R}^4$ and another point on the residual $\mathbb{CP}^1$. Therefore, the Euclidean twistor space is isomorphic to $\mathbb{R}^4\times\mathbb{CP}^1$, parametrized by $(x^{\dot\alpha\alpha},\lambda_\alpha)$.

A useful basis of the anti-holomorphic tangent and cotangent bundles of $\mathbb{CP}^{3}$ can be given in these coordinates
\begin{align}\label{basis}
\begin{split}
&\bar\partial_0 = \langle\lambda\hat\lambda\rangle\lambda^\alpha\frac{\partial}{\partial\hat\lambda^\alpha},\ \ \ \ \ \ \ \bar\partial_{\dot\alpha} = \lambda^\alpha\frac{\partial}{\partial x^{\dot\alpha\alpha}},\\
&\bar e^0 = \frac{\langle\hat\lambda d\hat\lambda\rangle}{\langle\lambda\hat\lambda\rangle^2},\ \ \ \ \ \ \ \ \ \ \ \bar e^{\dot\alpha} = \frac{\hat\lambda_\alpha \mathrm{d}x^{\dot\alpha\alpha}}{\langle\lambda\hat\lambda\rangle},
\end{split}
\end{align}
such that $\bar\partial=\bar e^0\bar\partial_0 + \bar e^{\dot\alpha}\bar\partial_{\dot\alpha}$ defines a complex structure on $\mathbb{CP}^3$. One can similarly define the basis $(\partial_0,\partial_{\dot\alpha})$ and $(e^0,e^{\dot\alpha})$ for the holomorphic bundles. Similar to the anti-holomorphic basis we choose a basis homogeneous in $\hat \lambda^\alpha$ as
\begin{equation}\label{basisHol}
    \begin{aligned}
        &\partial_0 = \frac{1}{\langle\lambda\hat\lambda\rangle}\hat\lambda^\alpha\frac{\partial}{\partial\lambda^\alpha},&\quad\partial_{\dot\alpha} &= \frac{1}{\langle\lambda\hat\lambda\rangle}\hat\lambda^\alpha\frac{\partial}{\partial x^{\dot\alpha\alpha}},\\
& e^0 = \langle\lambda d\lambda\rangle, & e^{\dot\alpha} &= \lambda_\alpha \mathrm{d}x^{\dot\alpha\alpha}.
    \end{aligned}
\end{equation}

\paragraph{Conformal structure}

The generators of the complexified conformal algebra can be represented by linear operators in the twistor space:
\begin{equation}
T^i{}_j = Z^i\frac{\partial}{\partial Z^j}
\end{equation}
which is a representation of $SL(4,\mathbb{C})$. These generators can be organized as:
\begin{align}
\begin{split}
&P_{\alpha\dot\alpha} = \lambda^\alpha\frac{\partial}{\partial\mu^{\dot\alpha}},\ \ \ J_{\alpha\beta} = \lambda_{(\alpha}\frac{\partial}{\partial\lambda^{\beta)}},\ \ \ \bar J_{\dot\alpha\dot\beta} = \mu_{(\dot\alpha}\frac{\partial}{\partial\mu^{\dot\beta)}}\\
&K^{\alpha\dot\alpha} = \mu^{\dot\alpha}\frac{\partial}{\partial \lambda_{\alpha}},\ \ \ D = \frac{1}{2}\left( \lambda_\alpha\frac{\partial}{\partial\lambda_\alpha} - \mu^{\dot\alpha}\frac{\partial}{\partial\mu^{\dot\alpha}} \right),
\end{split}
\end{align}
where $P_{\alpha\dot\alpha}, J_{\alpha\beta}, \bar J_{\dot\alpha\dot\beta}$ are the generators of translation and Lorentz symmetries, and $K^{\alpha\dot\alpha}$ are the generators of special conformal transformations (SCT), and $D$ is the dilatation generator.

\paragraph{Volume form}

We define the canonical holomorphic top form on $\mathbb{CP}^{3}$ as the following $(3,0)$-form:
\begin{equation}
\Omega^{3} = \epsilon_{ijkl}Z^idZ^j\wedge dZ^k\wedge dZ^l.
\end{equation}
During this work it will be useful to write this top form in terms of the basis (\ref{basisHol}) as:
\begin{equation}\label{eq:Omega3}
\Omega^3 = e^{\dot\alpha}\wedge e_{\dot\alpha}\wedge  e^{0}.
\end{equation}
Note that $\Omega^3$ is invariant under the $SL(4)$ symmetry corresponding to the conformal symmetries. Similarly, we can define the corresponding canonical anti-holomorphic top form:
\begin{equation}\label{eq:barOmega3}
\bar\Omega^3 = \epsilon_{ijkl}\bar Z^id\bar Z^j\wedge d\bar Z^k\wedge d\bar Z^l = \langle\lambda\hat\lambda\rangle^4\bar e^{\dot\alpha}\wedge \bar e_{\dot\alpha}\wedge\bar e^{0}.
\end{equation}
The volume form on twistor space is then defined as:
\begin{equation}
\text{vol}_{\mathbb{CP}^3} \equiv \frac{\Omega^3 \wedge \bar \Omega^3}{\langle\lambda\hat\lambda\rangle^4}=e^{\dot\alpha}\wedge e_{\dot\alpha}\wedge  e^{0}\wedge\bar e^{\dot\alpha}\wedge \bar e_{\dot\alpha}\wedge\bar e^{0},
\end{equation}
where the denominator ensures that the volume form is of homogeneous degree 0. Following the decomposition of the twistor space as $\mathbb{PT}^3\cong\mathbb{R}^4\times\mathbb{CP}^1$, this volume form also decomposes into spacetime and fibre components:
\begin{equation}\label{eq:SplitVol}
 \frac{\Omega^3 \wedge \bar \Omega^3}{\langle\lambda\hat\lambda\rangle^4} =  \mathrm{d}^4x\ \frac{\omega \wedge \bar \omega}{\langle\lambda\hat\lambda\rangle^2},
\end{equation}
where $\omega$ and $\bar\omega$ are the holomorphic and anti-holomorphic top from on the $\mathbb{CP}^1$-fibre:
\begin{align}
\begin{split}
&\omega = \langle\lambda\mathrm{d}\lambda\rangle\\
&\bar\omega = \langle\hat\lambda\mathrm{d}\hat\lambda\rangle
\end{split}
\end{align}
The volume form on the fibre $\mathbb{CP}^1$ is defined as:
\begin{equation}
\text{vol}_{\mathbb{CP}^1} \equiv \frac{\omega\wedge\bar\omega}{\langle\lambda\hat\lambda\rangle^2}=e^0\wedge\bar e^0.
\end{equation}
Equation \eqref{eq:SplitVol} thus shows that the volume form on twistor space $\mathbb{PT}^3$ decomposes into the product of the spacetime volume form and the volume form on the $\mathbb{CP}^1$-fibre.

\paragraph{Integration over the fibre} Throughout this work, we will be interested in reducing actions defined on twistor space $\mathbb{CP}^3$ to their corresponding actions on four-dimensional spacetime. To this end, we use the bundle structure of the twistor space and integrate over the $\mathbb{CP}^1$ fibre. After this integration, we obtain the effective corresponding theory on the four-dimensional spacetime. To show this in practice, consider an expression defined on $\mathbb{CP}^3$ of the form:
\begin{align}\label{eq:In}
I_n=\int_{\mathbb{CP}^3}\text{vol}_{\mathbb{CP}^3}\frac{\lambda^{\alpha_1}\cdots\lambda^{\alpha_n}\hat\lambda^{\beta_1}\cdots\hat\lambda^{\beta_n}}{\langle\lambda\hat\lambda\rangle^n}S_{\alpha_1\cdots\alpha_n}T_{\beta_1\cdots\beta_n},
\end{align}
where $S$ and $T$ are tensors depending on spacetime coordinates only. Using the decomposition in \eqref{eq:SplitVol}, we follow Appendix A of \cite{Mason} and use the following identity to integrate over $\mathbb{CP}^1$:
\begin{equation}
\int_{\mathbb{CP}^1} \text{vol}_{\mathbb{CP}^1}\frac{\lambda^{\alpha_1}\cdots\lambda^{\alpha_n}\hat\lambda^{\beta_1}\cdots\hat\lambda^{\beta_n}}{\langle\lambda\hat\lambda\rangle^n} =  -\frac{2\pi i}{(n+1)!}\sum_\sigma \epsilon^{\alpha_1\beta_{\sigma(1)}}\cdots\epsilon^{\alpha_n\beta_{\sigma(n)}}.
\end{equation}
Hence, the expression \eqref{eq:In}, originally defined on $\mathbb{CP}^3$, reduces to the spacetime expression
\begin{equation}\label{eq:IntegrateOut}
I_n = -\frac{2\pi i}{(n+1)}\int_{\mathbb{R}^4}\mathrm{d}^4x\ S_{\alpha_1\cdots\alpha_n}T^{\alpha_1\cdots\alpha_n}.
\end{equation}

\subsection{Twistor action and gauge theory}\label{subsec:twistorYM}
We first review the construction of gauge theory on $\mathbb{CP}^3$ and then show its equivalence to (perturbation theory around) the anti-self-dual Yang-Mills theory in four dimensional spacetime as we will re-derive the same relation for the twisted versions of these theories in the main part of this work in section \ref{sec:DefTwistor}. In particular, we will first introduce the so-called BF action, that is equivalent to the anti-self-dual sector of Yang-Mills followed by a discussion of a non-local term on twistor space, which completes the Yang-Mills theory in four dimensions. The equivalence between the twistor action and the spacetime Yang–Mills action is essentially demonstrated by reducing the theory from $\mathbb{CP}^3$ to $\mathbb{R}^4$. We will closely follow the construction of \cite{Mason}.

\paragraph{Anti-self-dual Yang-Mills theory from twistor space} To construct the desired gauge theory action on the twistor space, we need to consider a $(0,1)$-form $a$ of homogeneous degree 0, and another $(0,1)$-form $b$, which is inhomogeneous of degree $-4$. The action is defined as
\begin{align}\label{S1}
S_1[a,b] = &\frac{i}{2\pi}\int \Omega^3\wedge\tr\Big[b\wedge\big(\bar\partial a + a\wedge a\big)\Big],
\end{align}
where $\Omega^3$ is the canonical holomorphic measure defined in equation (\ref{eq:Omega3}). This action is invariant under the local gauge transformations
\begin{align}\label{gaugetransf}
\delta_\epsilon a = \bar\partial\epsilon + [\epsilon,a],\ \ \ \ \ \delta_\epsilon b = [\epsilon,b],
\end{align}
for $\epsilon$ being an infinitesimal gauge transformation, usually in $\mathfrak{su}(N)$ or $\mathfrak{u}(N)$. This action is homogeneous and invariant under conformal transformations. We now aim to show that this action is equivalent to the anti–self-dual Yang–Mills theory. To do so, we first use the gauge freedom (\ref{gaugetransf}) to impose a suitable gauge-fixing condition. In particular, we will choose a gauge that will fix the directions of the gauge fields along the residual $\mathbb{CP}^1$ directions. Once this gauge is chosen, the dependence on the fibre coordinates $(\lambda,\hat\lambda)$ becomes explicit, allowing the integration of the fibre. Performing this integration yields the corresponding four-dimensional action. The first step towards this procedure is to express the fields in the basis \eqref{basis} as
\begin{equation}
a = a_0\bar e^0 + a_{\dot\alpha}\bar e^{\dot\alpha},\ \ \ b = b_0\bar e^0 + b_{\dot\alpha}\bar e^{\dot\alpha}.
\end{equation}
We can then impose the gauge condition
\begin{equation}\label{eq:gaugeconditionYM}
\bar\partial^*|_X a_0=\bar\partial^*|_X b_0=0,
\end{equation}
for each $X\cong\mathbb{CP}^1$. The operator $\bar\partial^*|_X$ is the adjoint of the operator  $\bar\partial|_X = \bar e^0\bar\partial_0$. As $(0,1)$-forms on $X$, the fields $a_0$ and $b_0$ are automatically $\bar\partial$-closed, and then the gauge condition (\ref{eq:gaugeconditionYM}) enforces them to be harmonic along the fibres. By Hodge's theorem, the harmonic fields $a_0$ and $b_0$ lie in $H^1(\mathbb{CP}^1,{\cal O})$ and $H^1(\mathbb{CP}^1,{\cal O}(-4))$ respectively. Since $H^1(\mathbb{CP}^1,{\cal O})=0$, we can set $a_0=0$ as a gauge fixing, while $b_0$ is a (0,1)-form of a degree $-4$ and hence can be given as
\begin{align}\label{eq:gaugefixedYM}
a = a_{\dot\alpha}(x,\lambda,\hat\lambda)\bar e^{\dot\alpha},\ \ \ \ b = 3\frac{B_{\alpha\beta}(x)\hat\lambda^\alpha\hat\lambda^\beta}{\langle\lambda\hat\lambda\rangle^2}\bar e^0 +b_{\dot\alpha}(x,\lambda,\hat\lambda)\bar e^{\dot\alpha}.
\end{align}
Plugging these gauge-fixed expressions for the fields $a$ and $b$ back to the action (\ref{S1}), we get:
\begin{align}\label{S1two}
S_1[a,B]= \frac{i}{2\pi}\int \Omega^3\wedge\tr\Bigg[3\frac{B_{\alpha\beta}\hat\lambda^\alpha\hat\lambda^\beta}{\langle\lambda\hat\lambda\rangle^2}\bar e^0\wedge\left(\bar e^{\dot\alpha}\bar\partial_{\dot\alpha} a_{\dot\beta}\wedge\bar e^{\dot\beta} + a_{\dot\alpha}\bar e^{\dot\alpha}\wedge a_{\dot\beta}\bar e^{\dot\beta}\right) + b_{\dot\alpha}\bar e^{\dot\alpha}\wedge\Big(\bar e^{0}\bar\partial_0a_{\dot\beta}\wedge\bar e^{\dot\beta}\Big) \Bigg].
\end{align}
The component field $b_{\dot\alpha}$ appears only linearly in the action above and therefore plays the role of a Lagrange multiplier. Integrating out results in the equation
\begin{equation}
\bar e^0\bar\partial_0a_{\dot\alpha}=0 \Rightarrow a_{\dot\alpha}(x,\lambda,\hat\lambda) = a_{\dot\alpha}(x,\lambda).
\end{equation}
Since $a$ is of homogeneous degree 0 and $\bar e^{\dot\alpha}=\frac{\hat\lambda_\alpha}{\braket{\lambda\hat\lambda}} \mathrm{d} x^{\alpha\dot\alpha}$ of degree 1, it follows that
\begin{equation}\label{eq:FixingLambdaInA}
a_{\dot\alpha}(x,\lambda)=\lambda^\alpha A_{\alpha\dot\alpha}(x).
\end{equation}
Hence substituting (\ref{eq:FixingLambdaInA}) and (\ref{eq:gaugefixedYM}) in the action (\ref{S1two}) results in
\begin{align}
\begin{split}
S_1[a,b]=\ &\frac{i}{2\pi}\int \Omega\wedge\tr\Bigg[3\frac{B_{\alpha\beta}\hat\lambda^\alpha\hat\lambda^\beta}{\langle\lambda\hat\lambda\rangle^2}\bar e^0\wedge\left(\bar e^{\dot\alpha}\bar\partial_{\dot\alpha} a_{\dot\beta}\wedge\bar e^{\dot\beta} + a_{\dot\alpha}\bar e^{\dot\alpha}\wedge a_{\dot\beta}\bar e^{\dot\beta}\right) \Bigg]\\
=\ & \frac{i}{2\pi}\int \frac{\Omega\wedge\bar\Omega}{\langle\lambda\hat\lambda\rangle^4}\Bigg[3\frac{B_{\alpha\beta}\hat\lambda^\alpha\hat\lambda^\beta}{\langle\lambda\hat\lambda\rangle^2}\left(\bar\partial_{\dot\alpha} a_{\dot\beta}+ [a_{\dot\alpha},a_{\dot\beta}]\right)\epsilon^{\dot\alpha\dot\beta} \Bigg]\\
=\ & \frac{i}{2\pi}\int \frac{\Omega\wedge\bar\Omega}{\langle\lambda\hat\lambda\rangle^4}\Bigg[ 3B_{\gamma\delta}\left(\partial_{\alpha\dot\alpha} A_{\beta\dot\beta}+ [A_{\alpha\dot\alpha},A_{\beta\dot\beta}]\right)\epsilon^{\dot\alpha\dot\beta} \Bigg]\frac{\lambda^\alpha\lambda^\beta\hat\lambda^\gamma\hat\lambda^\delta}{\langle\lambda\hat\lambda\rangle^2},
\end{split}
\end{align}
where in the second line we used that the anti-holomorphic top form can be written as $\bar\Omega = \langle\lambda\hat\lambda\rangle^4\bar e^{\dot\alpha}\wedge\bar e_{\dot\alpha}\wedge\bar e^{0}$. Finally, integrating over the fibre by using the integration formula \eqref{eq:IntegrateOut}, one gets the four dimensional action:
\begin{equation}\label{eq:ASDYMindex}
S_1[a,b] = \int \mathrm{d}^4x\  \tr\left(B_{\alpha\beta}G^{\alpha\beta}\right),
\end{equation}
where $G_{\alpha\beta}$ is the anti-self-dual component of the field strength, with $A_{\alpha\dot\alpha}$ as the component of the gauge field:
\begin{equation}
G_{\alpha\beta} = (\partial_{\alpha\dot\alpha}A_{\beta\dot\beta} + [A_{\alpha\dot\alpha},A_{\beta\dot\beta}])\epsilon^{\dot\alpha\dot\beta}.
\end{equation}
The action (\ref{eq:ASDYMindex}) above describes the ASD Yang-Mills theory, as the equation of motion associated to the gauge field is:
\begin{equation}
G_{\alpha\beta}=0.
\end{equation}
In the context of twist deformations it is usefull to rewrite this action in terms of index free notation. For this purpose, we treat the gauge field $A_{\alpha\dot\alpha}$ as the component of a one-form:
\begin{equation}
A = A_{\alpha\dot\alpha}\mathrm{d}x^{\dot\alpha\alpha},
\end{equation}
and the ASD field strength and auxiliary field as the components of their corresponding ASD two-forms:
\begin{equation}\label{eq:GASDindexfree}
\begin{aligned}
G_{_{ASD}}&= G_{\alpha\beta}(\mathrm{d}x^{\dot\alpha\alpha}\wedge \mathrm{d}x^{\dot\beta\beta}\epsilon_{\dot\alpha\dot\beta})\\
B_{}=B_{ASD}&=B_{\alpha\beta}(\mathrm{d}x^{\dot\alpha\alpha}\wedge \mathrm{d}x^{\dot\beta\beta}\epsilon_{\dot\alpha\dot\beta}).
\end{aligned}
\end{equation}
The index-free action for the ASD Yang-Mills action then reads:
\begin{equation}\label{eq:ASDYMindexfree}
S_1[a,b]  = \int \tr\left(B_{}\wedge G_{_{ASD}}\right),
\end{equation}
\paragraph{Completing the Yang-Mills action} In the following we will review the construction of a non-local term that, once added to the action \eqref{eq:ASDYMindexfree}, will result in the full Yang-Mills theory as a perturbation around the anti-self-dual sector. In particular, we can add a topological term to the Yang-Mills action\cite{Adamo:Lectures}. The resulting action will reduce to
\begin{equation}\label{theta}
S_{\tiny \mbox{YM}-\theta} = \frac{1}{2g^2_{\tiny \mbox{YM}}}\int \tr\left( G_{_{ASD}}\wedge G_{_{ASD}} \right) ,
\end{equation}
which is perturbatively equivalent to the pure Yang-Mills action \cite{Adamo:Lectures}.\footnote{Note that in this action the parameter in front of the topological $\theta$ term is fixed. We will use the notation $S_{YM-\theta}$ in order to clarify that this action differs from the standard YM action by this particular choice of the $\theta$ term.} The latter can be expressed as a perturbation around the ASD Yang-Mills sector by introducing an anti-self-dual adjoint auxiliary field $B$, such that the action \eqref{theta} becomes
\begin{equation}\label{eq:BF4D}
S_{\tiny \mbox{YM}-\theta} = \int \tr \left( B_{}\wedge G_{_{ASD}} -\frac{g^2_{\tiny \mbox{YM}}}{2} B_{}\wedge B_{} \right).
\end{equation}
Note that setting $g_{YM}=0$ results in the action in \eqref{eq:ASDYMindexfree} leading to the anti-self-dual field equations for $G$ as being reproduced from the twistor action in \eqref{S1}.

In the remainder of this subsection, we will review how to reproduce the remaining part of \eqref{theta} from twistor space by following \cite{Mason}. The remaining term of the action \eqref{eq:BF4D} can be reproduced from twistor space by considering the nonlocal term \cite{Mason:twist}
\begin{equation}\label{S2}
S_2[b]= \frac{g^2_{\tiny \mbox{YM}}}{8\pi^2}\ \int_{\mathbb{R}^4} \mathrm{d}^4x\int_{\mathbb{CP}^1}\omega_1\int_{\mathbb{CP}^1}\omega_2\ \langle\lambda_1\lambda_2\rangle^2\ \tr\big(b_1b_2\big),
\end{equation}
where $b_i=b|_{X_i}$ where all the fibres are identified with each other as  $X_i=X$ and only differ by their parametrization $\lambda^\alpha_i$, such that on $X_i$, the incidence relation takes the form
\begin{equation}
    \mu_i^{\dot \alpha}=x^{\dot\alpha\alpha}\lambda^i_\alpha.
\end{equation}
It is easy to see that this action is invariant under the gauge transformation (\ref{gaugetransf}). We can therefore use the gauge fixing form of the field $b$, written in (\ref{eq:gaugefixedYM}), and substitute it in the action $S_2$ above. After integrating over the fibers, we get a four-dimensional action \cite{Mason, Mason:twist, Adamo:fish}:
\begin{equation}
S_2[b] = -\frac{g^2_{\tiny \mbox{YM}}}{2}\int \mathrm{d}^4x\ \tr(B_{\alpha\beta}B^{\alpha\beta}) = -\frac{g^2_{\tiny \mbox{YM}}}{2}\int \mathrm{d}^4x\ \tr\big(B_{}\wedge B_{}\big).
\end{equation}

\subsection{Super-twistor space}

We can extend the twistor space by adding ${\cal N}$ fermionic variables. The super-twistor space $\mathbb{PT}^{3|{\cal N}}$ is then defined as an open subset of $\mathbb{CP}^{3|{\cal N}}$, given by the open subset of $\mathbb{CP}^3$, together with $\mathcal{N}$ additional fermionic coordinates. As such, in homogeneous coordinates, the space $\mathbb{CP}^{3|\mathcal{N}}$ will be parametrized by $(Z^I,\psi^A)$. For the purpose of this work, we specify ${\cal N}=4$, since we aim to study the maximally super-symmetric Yang-Mills theory in four spacetime dimensions. The incidence relation is extended to the fermionic coordinates by
\begin{equation}
\psi^i = \theta^{i\alpha}\lambda_\alpha
\end{equation}
Similarly, we choose a reality condition for the fermionic variables as
\begin{equation}
\psi^i = (\psi^1,\psi^2,\psi^3,\psi^4) \mapsto \hat\psi^i = (-\bar\psi^2,\bar\psi^1,-\bar\psi^4,\bar\psi^3),
\end{equation}
which allows us to define the projection
\begin{equation}
\theta^{i\alpha} = \frac{\hat\psi^i\lambda^\alpha - \psi^i\hat\lambda^\alpha}{\langle\lambda\hat\lambda\rangle},
\end{equation}
hence the super-twistor space $\mathbb{PT}^{3|4}$ is isomorphic to $\mathbb{R}^{4|8}\times\mathbb{CP}^1$. The canonical holomorphic top form of the super-twistor space $\mathbb{CP}^{3|4}$ is defined as the following $(3|4,0)$-form:
\begin{equation}
\Omega^{3|4} = \epsilon_{ijkl}Z^idZ^j\wedge dZ^k\wedge dZ^l\wedge \mathrm{d}^4\psi = \Omega^3\wedge \mathrm{d}^4\psi
\end{equation}
and hence, the volume form is\footnote{Note that we are following the assumptions in \cite{Witten:CS}, that the fields on twistor space only depend on the holomorphic fermionic coordinates $\psi^I$ which is why the measure we consider does not include any integration over the anti-holomorphic coordinates $\hat \psi^I$.}
\begin{equation}
    \text{vol}_{\mathbb{CP}^{3|4}}=\frac{\Omega^{3|4}\wedge\bar{\Omega}^{3}}{\braket{\lambda\hat\lambda}^4}=\mathrm{d}x^{4|8}\wedge\frac{\omega\wedge\bar\omega}{\braket{\lambda\hat\lambda}^2},
\end{equation}
where
\begin{equation}
    \mathrm{d}^{4|8}x=\mathrm{d}^4xd^8\theta.
\end{equation}

\subsection{Supersymmetric gauge theory on super-twistor space}\label{subsec:twistorSYM}

Now, we will consider a supersymmetric extension of the BF theory discussed in section \ref{subsec:twistorYM}. For this purpose, we consider a $(0,1)$-form superfield $\mathcal{A}$, living on $\mathbb{CP}^{3|4}$, which can be expanded as:
\begin{align}\label{calA}
\begin{split}
{\cal A} = d\bar Z^{\bar I}\Big(&a_{\bar I}(Z,\bar Z) + \psi^i\tilde\chi_{\bar Ii} + \frac{1}{2!}\psi^i\psi^j\phi_{\bar I ij}(Z,\bar Z) + \frac{1}{3!}\epsilon_{ijkl}\psi^i\psi^j\psi^k\chi^l_{\bar I}(Z,\bar Z)\\
&\hspace{2cm} + \frac{1}{4!}\epsilon_{ijkl}\psi^i\psi^j\psi^k\psi^lb_{\bar I}(Z,\bar Z) \Big).
\end{split}
\end{align}
Note that the field $b$ from the BF theory in section \ref{subsec:twistorYM} becomes an auxiliary field in the superfield $\mathcal{A}$.
The superspace extension of the local action (\ref{S1}) of the previous section is given by a holomorphic Chern-Simons (hCS) action on $\mathbb{PT}^{3|4}$, which will be shown to describe the ASD sector of the ${\cal N}=4$ SYM theory. The action is given by:
\begin{equation}\label{6DCS}
S_{\tiny \mbox{hCS}}[{\cal A}] = \frac{i}{2\pi}\int_{\mathbb{CP}^{3|4}} \Omega^{3|4} \wedge \left({\cal A}\wedge\bar\partial{\cal A} + \frac{2}{3}{\cal A}\wedge{\cal A}\wedge{\cal A}\right),
\end{equation}
which is invariant under the gauge transformations
\begin{equation}\label{deltacalA}
\delta_\epsilon {\cal A} = \bar\partial\epsilon + [\epsilon,{\cal A}],
\end{equation}
for $\epsilon$ an element of the gauge algebra.

In order to obtain the anti-self-dual ${\cal N}=4$ SYM theory, we must partially gauge-fix the field ${\cal A}$ to remove the extra symmetry beyond the spacetime gauge group. We first expand ${\cal A}$ in the basis (\ref{basis}) as
\begin{equation}
{\cal A} = \bar e^0{\cal A}_0 + \bar e^{\dot\alpha}{\cal A}_{\dot\alpha},
\end{equation}
and impose the gauge condition
\begin{equation}
\bar\partial^*|_X{\cal A}_0 = 0,
\end{equation}
similar to the pure bosonic BF theory. Here, $\bar\partial^*|_X$ is the adjoint of the complex structure $\bar\partial$ restricted along a twistor line $X\cong\mathbb{CP}^1$. This is known as the \textit{Woodhouse} gauge condition \cite{Woodhouse}, and must hold for each component in the $\psi$-expansion separately,
\begin{equation}\label{eq:gaugeConditionYMComponents}
\bar\partial^*\big|_{X}a_0 = \bar\partial^*\big|_{X}\tilde\chi_{i,0} = \bar\partial^*\big|_{X}\phi_{ij,0} = \bar\partial^*\big|_{X}\chi^i_0 = \bar\partial^*\big|_{X}b_0=0.
\end{equation}
Since the fields are $(0,1)$-forms on $X$, they are automatically $\bar\partial$-closed. The gauge condition (\ref{eq:gaugeConditionYMComponents}) then implies that the component fields are harmonic along the fibres. This particularly fixes the components $a_0, \tilde\chi_{i,0}, \phi_{ij,0}, \chi^i_0$ and $b_0$, respectively and thus
\begin{align}\label{eq:gaugefixedN=4}
\begin{split}
a = a_{\dot\alpha}(x,\lambda,\hat\lambda)\bar e^{\dot\alpha},&\ \ \ \ \tilde\chi_i = \tilde\chi_{i\dot\alpha}(x,\lambda,\hat\lambda)\bar e^{\dot\alpha},\\
\phi_{ij} = \Phi_{ij}(x)\bar e^0+\phi_{ij\dot\alpha}(x,\lambda,\hat\lambda)\bar e^{\dot\alpha},&\ \ \ \ \chi^i = 2\frac{\Psi^{i\alpha}(x)\hat\lambda_\alpha}{\langle\lambda\hat\lambda\rangle}\bar e^0 + \chi^i_{\dot\alpha}(x,\lambda,\hat\lambda)\bar e^{\dot\alpha},\\
b = 3\frac{B_{\alpha\beta}(x)\hat\lambda^\alpha\hat\lambda^\beta}{\langle\lambda\hat\lambda\rangle^2}\bar e^0 & +b_{\dot\alpha}(x,\lambda,\hat\lambda)\bar e^{\dot\alpha}.
\end{split}
\end{align}
Plugging the gauge fixed fields into the action (\ref{6DCS}), we obtain:
\begin{align}
\begin{split}
S_{\tiny \mbox{hCS}_6}[{\cal A}] = \frac{i}{2\pi}\int \frac{\Omega\wedge\bar\Omega}{\langle\lambda\hat\lambda\rangle^4}\ \tr\Bigg[ &3\frac{B_{\alpha\beta}\hat\lambda^\alpha\hat\lambda^\beta}{\langle\lambda\hat\lambda\rangle^2}\left(\bar\partial_{\dot\gamma}a^{\dot\gamma} + \frac{1}{2}[a_{\dot\gamma},a^{\dot\gamma}]\right) + 2\frac{\Psi^{i\alpha}\lambda_\alpha}{\langle\lambda\hat\lambda\rangle}\left( \bar\partial_{\dot\gamma}\tilde\chi^{\dot\gamma}_i + [a_{\dot\gamma},\tilde\chi^{\dot\gamma}_i] \right) \\
&+\frac{1}{4}\epsilon^{ijkl}\Phi_{ij}\left(\bar\partial_{\dot\gamma}\phi_{kl}^{\dot\gamma} + [a_{\dot\gamma},\phi_{kl}^{\dot\gamma}] \right) + \frac{1}{2}\epsilon^{ijkl}\Phi_{ij}\tilde\chi_k^{\dot\gamma}\tilde\chi_{\dot\gamma l}\\
& + \Big(b^{\dot\alpha}\bar\partial_0a_{\dot\alpha} +\chi^{i\dot\alpha}\bar\partial_0\tilde\chi_{i\dot\alpha} + \frac{1}{2}\epsilon^{ijkl}\phi^{\dot\alpha}_{ij}\bar\partial_0\phi_{kl\dot\alpha} \Big)\Bigg].
\end{split}
\end{align}
The fields $b_{\dot\alpha}$ and $\chi^{i\dot\alpha}$ appear only linearly, and therefore, they play the role of Lagrange multipliers. Integrating them out enforces $\bar\partial_0a_{\dot\alpha} = \bar\partial_0\tilde\chi_{i\dot\alpha} =0$ and therefore, $a_{\dot\alpha}$ and $\tilde\chi_{i\dot\alpha}$ must be holomorphic in $\lambda$. Since they have homogeneous weights $+1$ and $0$ respectively, we find
\begin{equation}
a_{\dot\alpha}(x,\lambda,\hat\lambda) = \lambda^\alpha A_{\alpha\dot\alpha}(x),\ \ \ \ \tilde\chi_{i\dot\alpha}(x,\lambda,\hat\lambda) = \bar\Psi_{i\dot\alpha}(x)
\end{equation}
Similarly, because $\phi^{\dot\alpha}_{ij}$ appears only quadratically it may be eliminated using its equation of motion:
\begin{equation}
\bar\partial_0 \phi_{ij\dot\alpha} = \lambda^\alpha\left( \frac{\partial\Phi_{ij}}{\partial x^{\alpha\dot\alpha}} + [a_{\alpha\dot\alpha},\Phi_{ij}] \right) = 0,
\end{equation}
which is solved by:
\begin{equation}
\phi_{ij\dot\alpha} = \frac{\hat\lambda^\alpha}{\langle\lambda\hat\lambda\rangle}D_{\alpha\dot\alpha}\Phi_{ij}(x),
\end{equation}
where $D_{\alpha\dot\alpha}=\partial_{\alpha\dot\alpha}+A_{\alpha\dot\alpha}$ is the spacetime covariant derivative w.r.t. the gauge field $A_{\alpha\dot\alpha}$. Inserting the solutions for the fields $A_{\dot\alpha},\tilde\chi_{i\dot\alpha}$ and $\phi_{ij\dot\alpha}$ in the action results in
\begin{align}\label{SObarO}
\begin{split}
S_{\tiny \mbox{hCS}}[{\cal A}] = \frac{i}{2\pi}\int \frac{\Omega\wedge\bar\Omega}{\langle\lambda\hat\lambda\rangle^4}\tr\Bigg[ &3B_{\alpha\beta}G_{\gamma\delta}\frac{\lambda^\alpha\lambda^\beta\hat\lambda^\gamma\hat\lambda^\delta}{\langle\lambda\hat\lambda\rangle^2} + 2 \Psi_{i\alpha}D_{\beta\dot\beta}\bar\Psi^{i\dot\beta}\frac{\hat\lambda^\alpha\lambda^\beta}{\langle\lambda\hat\lambda\rangle}\\
& + \frac{1}{2}\epsilon^{ijkl}\Phi_{ij}D_{\alpha\dot\alpha}D^{\dot\alpha}{}_\beta\Phi_{kl}\frac{\hat\lambda^\alpha\lambda^\beta}{\langle\lambda\hat\lambda\rangle} + \frac{1}{2}\epsilon^{ijkl}\Phi_{ij}\bar\Psi_k^{\dot\gamma}\bar\Psi_{\dot\gamma l}\Bigg]
\end{split}
\end{align}
In the expression above, the dependence on the fibre coordinates $\lambda$ and $\hat\lambda$ is explicit. Hence, integrating over the fibre by using the integration formula \eqref{eq:IntegrateOut}, leads to the four dimensional action 
\begin{align}\label{eq:ASD-SYMAction}
S_{\tiny \mbox{hCS}}[{\cal A}] = \int \mathrm{d}^4x\ \tr\Bigg[ B_{\alpha\beta}G^{\alpha\beta} + \Psi^{i\alpha}D_{\alpha\dot\alpha}\bar\Psi_i^{\dot\alpha} - \frac{1}{4}D_{\alpha\dot\alpha}\Phi_{ij}D^{\alpha\dot\alpha}\Phi^{ij} + \bar\Psi_i^{\dot\alpha}[\Phi^{ij},\bar\Psi_{\dot\alpha j}]\Bigg],
\end{align}
which describes the anti-self-dual sector of the $\mathcal{N}=4$ SYM \cite{ASD-YM,Mason,Witten}. 

Like for the pure YM discussion in section \ref{subsec:twistorYM}, we add another nonlocal term to the twistor action, which will generate the non-anti-self-dual corrections to the $\mathcal{N}=4$ SYM. It is given by \cite{Mason:twist}
\begin{align}\label{eq:NonLocalUndeformedSYMExpanded}
\begin{split}
S_2[{\cal A}] =\ &-g^2_{\tiny \mbox{YM}}\oint d^{4|8}x \log{\det}\left(\bar\partial+{\cal A}\right)\big|_{\mathbb{CP}^1}\\
=\ &-g^2_{\tiny \mbox{YM}}\oint d^{4|8}x\ \tr\left(\log\bar\partial|_{\mathbb{CP}^1}\right)\\
&-g^2_{\tiny\mbox{YM}} \int d^{4|8}x\sum_{n=2}^\infty\frac{1}{n}\left(\frac{1}{2\pi i}\right)^n\int_{(\mathbb{CP})^n} \frac{\omega_1\cdots\omega_n}{\langle\lambda_1\lambda_2\rangle\cdots\langle\lambda_n\lambda_1\rangle}\tr\left({\cal A}_1\cdots{\cal A}_n\right).
\end{split}
\end{align}
This action is invariant under the gauge transformations (\ref{deltacalA}). In order to explicitly derive a four-dimensional action, the gauge field $\cal A$ is expressed in its component fields as an expansion in the fermionic coordinates $\psi^i$. To show the reduction to a four-dimensional action, we use the gauge fixing (\ref{eq:gaugefixedN=4}). This gauge fixing leaves only the $n=2,3,4$ terms in the infinite sum above. The reason is that the measure $d^{4|8}x$ involves an integration over $d^8\theta$, so only the terms proportional to $\theta^8$ contribute. Using the incidence relation $\psi^i = \theta^{i\alpha}\lambda_\alpha$, one see that we only need to keep the terms proportional to 8 instances of $\psi^i$. Moreover, in the action \eqref{eq:NonLocalUndeformedSYMExpanded}, the integration over the fibre ensures that only the zeroth components of the fields contribute, as these are 1-forms on $\mathbb{CP}^1$. For simplification, we incorporate the fermionic variables to the field components, and the superfield notation reads:
\begin{align}\label{componentsund}
&\tilde{\cal X}  = \tilde\chi_i\psi^i,\ \ \ \ \hat\Phi = \frac{1}{2}\phi_{ij}\psi^i\psi^j,\ \ \ \ {\cal X} = \frac{1}{3!}\chi^i\epsilon_{ijkl}\psi^j\psi^k\psi^l,\ \ \ \ {\cal B} = \frac{1}{4!}b\psi^4
\end{align}
This form of the fields allows us to write the contributions for the non-local terms more clearly:
\begin{equation}
S_2[{\cal A}] = S_2\Big|_{n=2} + S_2\Big|_{n=3} + S_2\Big|_{n=4},
\end{equation}
where:
\begin{align}
\begin{split}
S_2\big|_{n=2}=&-\frac{g^2_{\tiny\mbox{YM}}}{2}\left(\frac{1}{2\pi i}\right)^2\int d^{4|8}x \int_{(\mathbb{CP}^1)^2} \frac{\omega_1\omega_2}{\langle\lambda_1\lambda_2\rangle\langle\lambda_2\lambda_1\rangle}\tr({\cal B}_1{\cal B}_2),\\
S_2\big|_{n=3}=&-\frac{g^2_{\tiny\mbox{YM}}}{3}\left(\frac{1}{2\pi i}\right)^3 \int d^{4|8}x\int_{(\mathbb{CP}^1)^3} \frac{\omega_1\omega_2\omega_3}{\langle\lambda_1\lambda_2\rangle\langle\lambda_2\lambda_3\rangle\langle\lambda_3\lambda_1\rangle}\tr\left( {\cal X}_1 \hat\Phi_2 {\cal X}_3 \right), \\
S_2\big|_{n=4}= & -\frac{g^2_{\tiny\mbox{YM}}}{4}\left(\frac{1}{2\pi i}\right)^4\int d^{4|8}x \int_{(\mathbb{CP}^1)^4}\frac{\omega_1\omega_2\omega_3\omega_4}{\langle\lambda_1\lambda_2\rangle \langle\lambda_2\lambda_3\rangle \langle\lambda_3\lambda_4\rangle \langle\lambda_4\lambda_1\rangle}\tr\left( \hat\Phi_1\hat\Phi_2\hat\Phi_3\hat\Phi_4 \right).
\end{split}
\end{align}
In principle, the $n=3$ case contains a contribution of the form ${\cal B}\hat\Phi\hat\Phi$. However, this term can be neglected since this does not result in a non-vanishing scalar after integrating over the fibre. Specifically, the integration will lead to a term proportional to $B_{\alpha\beta}\epsilon^{\alpha\beta}=0$. The only non-vanishing contributions to the integrals of the non-local terms come from the zeroth components of the fields.

The $n=2$ term corresponds to the pure Yang-Mills contribution discussed earlier. To see this, we make use of the identity \cite{Mason}:
\begin{equation}
\int d^8\theta\  (\psi_1)^4(\psi_2)^4\Big|_{\psi^i=\theta^{i\alpha}\lambda_\alpha} = \langle\lambda_1\lambda_2\rangle^4,
\end{equation}
which collapses the Grassmann integral to the familiar spinor bracket. Applying this identity reduces the action $S_2\big|_{n=2}$ above to the non-local action (\ref{S2}), and thus reproduces the part of the twisted YM theory that complements the ASD sector discussed in the previous section:
\begin{equation}
S_2\big|_{n=2}=-\frac{g^2_{\tiny \mbox{YM}}}{2} \int d^{4}x\ \tr\big( B_{\alpha\beta}B^{\alpha\beta}\big)
\end{equation}
The $n=3$ case gives the complement of the interaction between scalar bosons and fermions, which is of the form $\Psi\Phi\Psi$:
\begin{equation}\label{n=3}
\begin{aligned}
S_2\Big|_{n=3}=-\frac{g^2_{\tiny\mbox{YM}}}{3(2\pi i)^3}\int d^{4|8}x &\int \frac{\omega_1\omega_2\omega_3}{\langle\lambda_1\lambda_2\rangle\langle\lambda_2\lambda_3\rangle\langle\lambda_3\lambda_1\rangle}\times\\
&\times\tr\left(  (\psi^i_1\psi^j_1\psi^k_1)\Psi^l_0\epsilon_{ijkl}(\psi^m_2\psi^n_2)\Phi_{0,mn}(\psi^p_3\psi^q_3\psi^r_3)\Psi^s_0\epsilon_{pqrs}  \right).
\end{aligned}
\end{equation}
Integrating over the fermionic components and substituting the gauge-fixed expressions for the fields $\Phi_0$ and $\Psi_0$ (equation \ref{eq:gaugefixedN=4}) to integrate out the $\mathbb{CP}^1$-fibres, we obtain for the $n=3$ contribution:
\begin{align}
\begin{split}
S_2\Big|_{n=3}=&\ -\frac{g^2_{\tiny\mbox{YM}}}{3(2\pi i)^3}\int \mathrm{d}^4x \int \frac{\omega_1\wedge\bar\omega_1}{\langle\lambda_1\hat\lambda_1\rangle^2}\int \frac{\omega_2\wedge\bar\omega_2}{\langle\lambda_2\hat\lambda_2\rangle^2} \int \frac{\omega_3\wedge\bar\omega_3}{\langle\lambda_3\hat\lambda_3\rangle^2}\ \tr\left(\frac{\hat\lambda_1^\alpha}{\langle\lambda_1\hat\lambda_1\rangle}\Psi^{i}_\alpha  \Phi_{ij} \frac{\hat\lambda_3^\beta}{\langle\lambda_3\hat\lambda_3\rangle} \Psi_\beta^{j} \right)\langle\lambda_1\lambda_3\rangle\\
= &\ -g^2_{\tiny \mbox{YM}}\int \mathrm{d}^4x\ \tr\Big(\Psi^i_\alpha[\Phi_{ij},\Psi^{j\alpha}]\Big),
\end{split}
\end{align}
which reproduces the missing Yukawa interactions for $\mathcal{N}=4$ SYM. The last non-trivial term is the $n=4$ case, and will give the quartic interaction between the scalar bosons:
\begin{equation}\label{n=4}
\begin{aligned}
S_2\Big|_{n=4}=-\frac{g^2_{\tiny\mbox{YM}}}{4(2\pi i)^4}\int d^{4|8}x &\int \frac{\omega_1\cdots\omega_4}{\langle\lambda_1\lambda_2\rangle\cdots\langle\lambda_4\lambda_1\rangle}\times\\
&\times\tr\left(  (\psi^i_1\psi^j_1)\Phi_{0,ij} (\psi^k_2\psi^l_2)\Phi_{0,kl} (\psi^m_3\psi^n_3)\Phi_{0,mn} (\psi^p_4\psi^q_4)\Phi_{0,pq} \right).
\end{aligned}
\end{equation}
We now use the result of \cite{Koster} section 2.5 to integrate over the $\mathbb{CP}^1$ fibres. The expression above then reduces to:
\begin{equation}
S_2\Big|_{n=4} =  \frac{g^2_{\tiny\mbox{YM}}}{4}\int \mathrm{d}^4x\ \tr\Big([\Phi^{ij},\Phi^{kl}][\Phi_{ij},\Phi_{kl}]\Big).
\end{equation}
Summing the three terms, we conclude that the non-local term is then given by:
\begin{align}
\begin{split}
S_2[{\cal A}] = &\ S_2\Big|_{n=2} + S_2\Big|_{n=3} + S_2\Big|_{n=4}\\
= &\ g^2_{\tiny\mbox{YM}}\int \mathrm{d}^4x\ \tr\left( -\frac{1}{2}B_{\alpha\beta}B^{\alpha\beta} -\Psi^i_\alpha[\Phi_{ij},\Psi^{j\alpha}] + \frac{1}{4}[\Phi^{ij},\Phi^{kl}][\Phi_{ij},\Phi_{kl}]  \right).
\end{split}
\end{align}
The sum of the two actions $S_{\tiny \mbox{hCS}}+S_2$ gives: 
\begin{align}\label{eq:SYMactionindex}
\begin{split}
S_{\tiny \mbox{hCS}} + S_2 = S_{\tiny \mbox{SYM}}&=\int \mathrm{d}^4x\ \tr\left(B_{\alpha\beta} G^{\alpha\beta} - \frac{g^2_{\tiny\mbox{YM}}}{2}B_{\alpha\beta}B^{\alpha\beta}\right)\\
&\ \ \ -\frac{1}{4} \int \mathrm{d}^4x\ \tr\left(D_{\alpha\dot\alpha}\Phi^{ij}D^{\alpha\dot\alpha}\Phi_{ij}\right) + \int \mathrm{d}^4x\   \tr\Big(\Psi^{i\alpha}D_{\alpha\dot\alpha}\bar\Psi^{\dot\alpha}_i \Big)\\
&\ \ \ + \int \mathrm{d}^4x\ \tr\Big( \bar\Psi_i^{\dot\alpha}  [\Phi^{ij},\bar\Psi_{j\dot\alpha}] \Big) -g^2_{\tiny\mbox{YM}} \int \mathrm{d}^4x\ \tr\Big(\Psi^{i\alpha}[\Phi_{ij},\Psi^j_{\alpha}]\Big)\\
&\ \ \ +\frac{g^2_{\tiny\mbox{YM}}}{4}  \int \mathrm{d}^4x\ \tr\Big( [\Phi^{ij},\Phi^{kl}][\Phi_{ij},\Phi_{kl}]\Big),
\end{split}
\end{align}
which after rescaling $\bar\Psi\to \sqrt{g_{\tiny\mbox{YM}}}\bar\Psi$ and $\Psi\to \Psi/\sqrt{g_{\tiny\mbox{YM}}}$, integrating out the auxiliary field $B$ yields the complete ${\cal N} = 4$ SYM action up to a topological $\theta$-term, $\int \tr(G\wedge G)$, that does not contribute perturbatively.

\section{Twist-noncommutative Gauge theory}\label{sec:DefBFTheory}

In the context of noncommutative gauge theories --particularly Yang-Mills theory and its supersymmetric extension, ${\cal N}=4$ SYM -- recent developments led to a consistent gauge invariant construction of an action in the setting of noncommutativities generated by a twist of the underlying Hopf algebra of Poincar\'e symmetries of the theory \cite{Meier:quad,Meier:twist}. In this section, we will review this construction.

\subsection{Drinfel'd twists and twisted Hopf algebras}
\label{sec:twist}
The Poincar\'e algebra $\mathcal{P}$ acts naturally on fields via Lie derivatives w.r.t. the vector field representation of the Poincar\'e algebra. This Lie algebra is extended to a Hopf algebra in order to act on products of fields. This is achieved by introducing the standard coproduct, counit and antipode on the universal enveloping algebra $\mathcal{U}(\mathcal{P})$ of the Poincar\'e algebra
\begin{align*}
\Delta(X)&=1\otimes X+X\otimes1&\Delta(1)&=1\otimes1\\
S(X)&=-X&S(1)&=1\\
\varepsilon(X)&=0&\varepsilon(1)&=1,
\end{align*}
for $x\in \mathcal{P}$, extended to $\mathcal{U(P)}$. 

\paragraph{Drinfel'd twist and deformed Hopf algebra} A Drinfel'd twist $\mathcal{F}$ defines a deformation of the Hopf algebra structure above. It is an invertible element of $\mathcal{U(P)}\otimes \mathcal{U(P)}$, expandable in a deformation parameter satisfying the cocycle and normalization conditions
\begin{equation}
\begin{aligned}
\label{eq:cocycle}
\left(\mathcal{F}\otimes1\right)\left(\Delta\otimes1\right)\mathcal{F}&=\left(1\otimes\mathcal{F}\right)\left(1\otimes\Delta\right)\mathcal{F}\\
\left(1\otimes\varepsilon\right)\mathcal{F}=(\varepsilon&\otimes1)\mathcal{F}=1\otimes1.
\end{aligned}
\end{equation}
Furthermore, it will be useful to express the Drinfel'd twist and its inverse as an implicit sum over tensor products between distinct elements $f^\alpha,f_\alpha,\bar f^\alpha,\bar f_\alpha\in \mathcal{U(P)}$ as
\begin{equation}
    \begin{aligned}
        \mathcal{F}&=f^\alpha\otimes f_\alpha & \qquad \mathcal{F}^{-1}=\bar{\mathcal{F}}=\bar f^\alpha \otimes \bar f_\alpha.
    \end{aligned}
\end{equation}

The latter twist can be used to generate a deformation of the Hopf algebra above by defining a new coproduct and antipode as
\begin{equation}
    \begin{aligned}
        \Delta_\mathcal{F}(X)&=\mathcal{F}\Delta(X)\bar{\mathcal{F}}\\
        S_\mathcal{F}(X)&=f^\alpha S(f_\alpha) S(X) S(\bar{f}^\beta)\bar{f}_\beta.
    \end{aligned}
\end{equation}

To have a representation of fields compatible with the latter twisted Hopf algebra, the product of fields is replaced by a so-called star product:
\begin{equation}
    (f\star g)(x)=\mu_\mathcal{F}(f\otimes g)(x)=\mu(\bar{\mathcal{F}}(f\otimes g))(x),
\end{equation}
where $\mu(f\otimes g)(x)=f(x)g(x)$ is the pointwise product between fields. The star product is noncommutative but remains associative. The noncommutativity is captured by the R matrix $\mathcal{R}=\mathcal{F}_{op}\bar{\mathcal{F}}=R^\alpha\otimes R_\alpha$, such that
\begin{equation}
    f\star g=\mu (\bar{\mathcal{F}}\mathcal{R}_{op}(g\otimes f)).
\end{equation}
Furthermore, $R$ satisfies the Yang-Baxter equation as a result of the cocycle condition for the twist and defines a triangular Hopf algebra as $\mathcal{R}_{op}=\bar{\mathcal{R}}$. To leading order, $\mathcal{R}=1\otimes 1+\lambda r$. Hence the noncommutativity of coordinates is captured by
\begin{equation}
    \scom{x^\mu}{x^\nu}=x^\mu\star x^\nu-x^\nu\star x^\mu=\lambda\mu(r(x^\mu\otimes x^\nu))+\mathcal{O}(\lambda^2)=\lambda r^{\mu\nu}+\mathcal{O}(\lambda^2),
\end{equation}
where $r=\frac{1}{2}r^{\mu\nu}\partial_\mu\wedge\partial_\nu$. In the context of noncommutative field theories the components $\lambda r^{\mu\nu}$ are also denoted as $\theta^{\mu\nu}$.

\subsection{Twisted differential calculus}\label{sec:DiffCalc}
In order to define field theory actions, in particular for gauge theories on the twisted spacetime, a twisted differential calculus is needed. The twisted differential calculus reviewed in this section has been previously discussed in \cite{Aschieri:2005zs,Aschieri:2009ky}. In \cite{Meier:twist}, a consistent star-linear Hodge duality has been defined by keeping track of the appearance of the twist and the R matrix in various representations. In the following, we will follow the presentation of \cite{Meier:twist}. 
\paragraph{Exterior algebra of differential forms} Similarly to the star product between functions, the wedge product between arbitrary forms $\omega$ and $\xi$ is twisted as
\begin{equation}
    \omega\wedge_\star\xi=\bigwedge\left( \bar{\mathcal{F}} (\omega\otimes\xi) \right)=\bar f^\alpha\omega\wedge \bar f_\alpha\xi.
\end{equation}
Similarly to the algebra of functions in section \ref{sec:twist}, the algebra of differential forms with the deformed product $\wedge_\star$ forms a representation of the twisted Hopf algebra. 
\paragraph{Exterior derivative} Beyond the exterior algebra, we need an exterior derivative. The conventional exterior derivative by construction commutes with Lie derivatives and hence with the action of the twist on arbitrary differential forms. Hence, it satisfies the usual Leibniz rule
\begin{equation}
    \form{\left( \omega \wedge_\star \xi \right)}=\form{\omega}\wedge_\star \xi+ (-1)^{\abs{\omega}}\omega\wedge_\star\form{\xi},
\end{equation}
which will serve as the exterior derivative in the deformed setting as well.

For a given basis of one-form , we can choose to express an arbitrary differential form w.r.t. the twisted wedge product of forms or the undeformed wedge product, i.e. for a one-form $A$
\begin{equation}
    A=A_\mu\form{x^\mu}=A^\star_\mu\star \form{x^\mu}.
\end{equation}
As the twist acts as Lie derivatives, acting on a basis one-form  it picks up a pair of indices \cite{Meier:twist}. Hence,
\begin{equation}
\begin{aligned}
    \form{x^\mu}\star f(x) &= \form{x^\nu}\tensor{\bar F}{_\nu^\mu}f(x)\\
    \form{x^\mu}\star f(x) &= \left(\tensor{R}{_\nu^\mu}f(x)\right)\star\form{x^\nu}\\
    \form{x^\mu}\wedge_\star\form{x^\nu} &= -\tensor{R}{_\rho^\mu_\sigma^\nu}\form{x^\sigma}\wedge_\star\form{x^\rho},
\end{aligned}    
\end{equation}
where $\tensor{\bar F}{_\nu^\mu}$ and $\tensor{R}{_\nu^\mu}$ still act as a differential operator on the function moved. As the twist is built from Poincar\'e generators, $R$ and $F$ as matrices are operator-valued Poincar\'e transformation matrices. Hence,
\begin{equation}
\begin{aligned}
    \tensor{\bar R}{_\mu^\nu}&=\tensor{R}{^\nu_\mu}&\qquad&&\tensor{\bar F}{_\mu^\nu}&=\tensor{F}{^\nu_\mu} 
\end{aligned}
\end{equation}

\paragraph{Integral cyclicity and unimodularity} As the star product is noncommutative, the star-wedge product is not graded antisymmetric. In \cite{Meier:twist}, it was argued however, that graded cyclicity is desirable for a gauge invariant theory. This is guaranteed by twists satisfying \cite{Aschieri:2009ky}
\begin{equation}\label{eq:UnimodCondition}
S(\bar{f}^\alpha)\bar{f}_\alpha=1.
\end{equation}

\paragraph{Hodge duality} For a formulation of noncommutative Yang Mills theory, the discussion on the twisted differential calculus needs to be completed by a formulation of Hodge duality and the Hodge-star operator in the twisted setting. In fact, in the setup of twist deformations of the Poincar\'e algebra, this has been constructed in \cite{Meier:twist,Meier:quad} and a similar discussion for q-Minkowski space in \cite{Majid:1994mh,Meyer:1994wi}.

First, a totally R-antisymmetric generalized Levi-Civita symbol is defined as
\begin{equation}
   \epsilon_\star^{\mu\nu\rho\sigma}\form{x^0}\wedge_\star\form{x^1}\wedge_\star\form{x^2}\wedge_\star\form{x^3}=\form{x^\mu}\wedge_\star\form{x^\nu}\wedge_\star\form{x^\rho}\wedge_\star\form{x^\sigma},
\end{equation}
which for twists in the Poincar\'e algebra is star commutative. In order to distinguish the deformed Levi-Civita symbol from its undeformed counterpart, we put stars between the indices. Note that top forms are graded cyclic, which effectively allows us to remove one star product, i.e.:
\begin{equation}
\begin{aligned}
\form{x^\mu}\wedge_\star\form{x^\nu}\wedge_\star\form{x^\rho}\wedge_\star\form{x^\sigma}&=\left(\form{x^\mu}\wedge_\star\form{x^\nu}\right)\wedge\left(\form{x^\rho}\wedge_\star\form{x^\sigma}\right)
\end{aligned}
\end{equation}
We can furthermore express the remaining star products explicitly using the vector representation of the twist and hence
\begin{equation}
    \epsilon_\star^{\mu\nu\rho\sigma}=\tensor{\bar F}{_\theta^\mu_\lambda^\nu}\tensor{\bar F}{_\gamma^\rho_\delta^\sigma}\epsilon^{\theta\lambda\gamma\delta}.
\end{equation}

The deformed Levi-Civita symbol is now used to define a Hodge duality on basis forms as \cite{Meier:twist}
\begin{equation}
    *\form{x^{\mu_1}}\wedge_\star...\wedge_\star\form{x^{\mu_{k}}}=\frac{(-1)^{\sigma(k)}}{(4-k)!}\tensor{\epsilon}{^\star_{\mu_{k+1}...\mu_4}^{\mu_1...\mu_k}}\form{x^{\mu_4}}\wedge_\star...\wedge_\star\form{x^{\mu_{k+1}}}.
\end{equation}
For deformations in the Poincar\'e algebra this is both linear and star linear\cite{Meier:twist}, which allows to extend the latter definition to arbitrary differential forms.

\paragraph{Twisted spinor index basis}
In order to make contact to twistor space and BF theory, which naturally describes SD and ASD YM sectors, we would like to introduce a basis of self-dual and anti-self-dual two forms on twist deformed spacetime. For this purpose, we introduce the standard spinor index basis for a given vector $v^\mu$, i.e.
\begin{equation}
    v^{\dot\alpha\alpha}=\frac{1}{\sqrt{2}}\sigma^{\dot\alpha\alpha}_\mu v^\mu
\end{equation}
and hence
\begin{equation}
    \form{x^{\dot\alpha\alpha}}=\frac{1}{\sqrt{2}}\sigma^{\dot\alpha\alpha}_\mu \form{x^\mu}.
\end{equation}
As the Pauli matrices convert a Poincar\'e transformation into its left- and right-handed spinor representation and $\tensor{R}{_\mu^\nu}$ and $\tensor{F}{_\mu^\nu}$ are Poincar\'e transformations acting on a basis form, it follows \cite{Meier:twist}
\begin{equation}
\label{eq:HandedRMatRep}
\begin{aligned}
    \tensor{R}{_\mu^\nu}\sigma_\nu^{\dot\alpha\alpha}&=\sigma_\mu^{\dot\beta\beta}\tensor{R}{_\beta^\alpha}\tensor{R}{_{\dot\beta}^{\dot\alpha}}\\
    \tensor{F}{_\mu^\nu}\sigma_\nu^{\dot\alpha\alpha}&=\sigma_\mu^{\dot\beta\beta}\tensor{F}{_\beta^\alpha}\tensor{F}{_{\dot\beta}^{\dot\alpha}},
\end{aligned}
\end{equation}
where $\tensor{R}{_\beta^\alpha}$ and $\tensor{F}{_\beta^\alpha}$ are the left-handed part and $\tensor{R}{_{\dot\beta}^{\dot\alpha}}$ and $\tensor{F}{_{\dot\beta}^{\dot\alpha}}$ the right-handed part of the Poincar\'e transformations $R$ and $F$, respectively. In the of twistor space, the antisymmetric tensor $\epsilon$ is an invariant of the ${\cal R}$ matrix, which means in coordinates:
\begin{equation}\label{RR}
\begin{aligned}
&\epsilon_{\gamma\delta} \tensor{R}{_\alpha^\gamma} \tensor{R}{_\beta^\delta} = \epsilon_{\alpha\beta}\\
&\epsilon_{\dot\gamma\dot\delta} \tensor{R}{_{\dot\alpha}^{\dot\gamma}} \tensor{R}{_{\dot\beta}^{\dot\delta}} = \epsilon_{\dot\alpha\dot\beta}
\end{aligned}
\end{equation}
A light-like vector $v_{null}$ decomposes into the product of a left-handed and a right-handed spinor $a^\alpha$ and $a^{\dot \alpha}$
\begin{equation}
    v_{null}^{\dot \alpha \alpha}=a^{\dot \alpha}a^\alpha.
\end{equation}
This structure, however is written in terms of the undeformed product. In the nomcommutative setting we rather want to have a structure, which is compatible with the star product. Hence, we define the noncommutative equivalent of $v_{null}^{\dot \alpha \alpha}$ as
\begin{equation}
    v_{null}^{\dot \alpha \star \alpha}=\tensor{F}{_{\dot \beta}^{\dot\alpha}_{\beta}^\alpha}a^{\dot\beta} a^{\beta}.
\end{equation}
Note that $v_{null}^{\dot \alpha \alpha}$ is a null vector while $v_{null}^{\dot \alpha \star \alpha}$ is the noncommutative version associated to the null vector $v_{null}^{\dot \alpha \alpha}$ but is not a null vector by itself in general. As a consequence, it is consistent to define
\begin{equation}
    v^{\dot \alpha \star \alpha}=\frac{1}{\sqrt{2}}\tensor{\bar F}{_{\dot \beta}^{\dot\alpha}_{\beta}^\alpha}\sigma^{\dot\beta\beta}_\mu v^\mu
\end{equation}
for any vector $v^\mu$. Within the deformed spinor basis, we can choose a basis of one-form  as
\begin{equation}
\label{eq:SpinorBas2Form}
    \begin{aligned}
        \mathrm{d}x^{\dot \alpha \star \alpha}=\frac{1}{\sqrt{2}}\tensor{\bar F}{_{\dot \beta}^{\dot\alpha}_{\beta}^\alpha}\sigma^{\dot\beta\beta}_\mu \mathrm{d}x^\mu.
    \end{aligned}
\end{equation}

\paragraph{(Anti-)self-dual two-forms}
As in the undeformed case, we would like to choose a pure chiral basis built from pure left-handed and right-handed two-forms. In this direction, we can show that the undeformed chiral basis in fact remains an eigenbasis of the deformed Hodge operator. For example,
\begin{equation}\label{eq:AntiSelfDualUndeformed}
    \begin{aligned}
        *\left( \mathrm{d}x^{\dot\alpha\alpha}\wedge\mathrm{d}x^{\dot\beta \beta}\epsilon_{\alpha\beta} \right)&=\tensor{F}{_{\dot\gamma}^{\dot\alpha}_{\delta}^{\beta}}\tensor{F}{_{\gamma}^{\alpha}_{\dot\delta}^{\dot\beta}} \epsilon_{\alpha\beta}*\left( \mathrm{d}x^{\dot\gamma\gamma}\wedge_\star\mathrm{d}x^{\dot\delta \delta} \right)\\
        &=\frac{1}{2}\tensor{F}{_{\dot\gamma}^{\dot\alpha}_{\delta}^{\beta}}\tensor{F}{_{\gamma}^{\alpha}_{\dot\delta}^{\dot\beta}} \epsilon_{\alpha\beta} \sigma_\mu^{\dot\gamma\gamma}\sigma_\nu^{\dot\delta\delta}*\left( \mathrm{d}x^\mu\wedge_\star\mathrm{d}x^\nu \right)\\
        &=\frac{1}{4}\tensor{F}{_\mu^\rho_\nu^\sigma}\sigma_\rho^{\dot\alpha\alpha}\sigma_\sigma^{\dot\beta\beta}\epsilon_{\alpha\beta}\epsilon_\star^{\tau\kappa\mu\nu}\mathrm{d}x^{\kappa}\wedge_\star\mathrm{d}x^{\tau}\\
        &=\frac{1}{4}\sigma_\mu^{\dot\alpha\alpha}\sigma_\nu^{\dot\beta\beta}\epsilon_{\alpha\beta}\epsilon^{\tau\kappa\mu\nu}\mathrm{d}x^{\kappa}\wedge\mathrm{d}x^{\tau}\\
        &=-\mathrm{d}x^{\dot\alpha\alpha}\wedge\mathrm{d}x^{\dot\beta \beta}\epsilon_{\alpha\beta}
    \end{aligned}
\end{equation}
and similarly for $*\left( \mathrm{d}x^{\dot\alpha\alpha}\wedge\mathrm{d}x^{\dot\beta \beta}\epsilon_{\dot\alpha\dot\beta} \right)=\mathrm{d}x^{\dot\alpha\alpha}\wedge\mathrm{d}x^{\dot\beta \beta}\epsilon_{\dot\alpha\dot\beta}$. We can then express every star-basis two-form in terms of the latter chiral basis as follows
\begin{equation}
\begin{aligned}
    \mathrm{d}x^{\dot\alpha\star\alpha}\wedge_\star\mathrm{d}x^{\dot\beta\star\beta}&=\tensor{\bar F}{_{\dot\gamma}^{\dot\gamma'}_{\gamma}^{\gamma'}}\tensor{\bar F}{_{\dot\gamma'}^{\dot\alpha}_{\delta}^{\delta'}}\tensor{\bar F}{_{\gamma'}^{\alpha}_{\dot\delta}^{\dot\delta'}}\tensor{\bar F}{_{\dot\delta'}^{\dot\beta}_{\delta'}^{\beta}}\mathrm{d}x^{\dot\gamma\gamma}\wedge\mathrm{d}x^{\dot\delta\delta}\\
    &=\tensor{\bar F}{_{\dot\gamma}^{\dot\gamma'}_{\gamma}^{\gamma'}}\tensor{\bar F}{_{\dot\gamma'}^{\dot\alpha}_{\delta}^{\delta'}}\tensor{\bar F}{_{\gamma'}^{\alpha}_{\dot\delta}^{\dot\delta'}}\tensor{\bar F}{_{\dot\delta'}^{\dot\beta}_{\delta'}^{\beta}}\\
    &\times\left(\epsilon^{\gamma\delta}\mathrm{d}x^{\dot\gamma\tau}\wedge\mathrm{d}x^{\dot\delta\kappa}\epsilon_{\tau\kappa}+ \epsilon^{\dot\gamma\dot\delta}\mathrm{d}x^{\dot\tau\gamma}\wedge\mathrm{d}x^{\dot\kappa\delta}\epsilon_{\dot\tau\dot\kappa}\right).
\end{aligned}
\end{equation}
We can simplify this expression by using
\begin{equation}
\begin{aligned}
    \tensor{\bar F}{_{\dot\gamma}^{\dot\gamma'}_{\gamma}^{\gamma'}}\tensor{\bar F}{_{\dot\gamma'}^{\dot\alpha}_{\delta}^{\delta'}}\tensor{\bar F}{_{\gamma'}^{\alpha}_{\dot\delta}^{\dot\delta'}}\tensor{\bar F}{_{\dot\delta'}^{\dot\beta}_{\delta'}^{\beta}}\epsilon^{\gamma\delta}
    &= \tensor{ F}{_{\dot\gamma}^{\dot\gamma'}_{\gamma}^{\delta}}\tensor{\bar F}{_{\dot\gamma'}^{\dot\alpha}_{\delta}^{\delta'}}\tensor{\bar F}{_{\gamma'}^{\alpha}_{\dot\delta}^{\dot\delta'}}\tensor{\bar F}{_{\dot\delta'}^{\dot\beta}_{\delta'}^{\beta}}\epsilon^{\gamma\gamma'}\\
   &=\tensor{\bar F}{_{\gamma'}^{\alpha}_{\dot\delta}^{\dot\delta'}}\tensor{\bar F}{_{\dot\delta'}^{\dot\beta}_{\gamma}^{\beta}}\epsilon^{\gamma\gamma'}\delta_{\dot\gamma}^{\dot\alpha}\\
   &=\tensor{\bar F}{_{\gamma'}^{\alpha}_{\dot\delta}^{\dot\delta'}}\tensor{\left.F_{op}\right.}{_{\gamma}^{\gamma'}_{\dot\delta'}^{\dot\beta}}\epsilon^{\gamma\beta}\delta_{\dot\gamma}^{\dot\alpha}\\
   &=\tensor{R}{_\gamma^\alpha_{\dot\delta}^{\dot\beta}}\epsilon^{\gamma\beta}\delta_{\dot\gamma}^{\dot\alpha}
    \end{aligned}
\end{equation}
and similarly for the second term, which then leads to
\begin{equation}\label{eq:PureChiralBasisDeformed}
    \mathrm{d}x^{\dot\alpha\star\alpha}\wedge_\star\mathrm{d}x^{\dot\beta\star\beta}=\tensor{R}{_\gamma^\alpha_{\dot\gamma}^{\dot\beta}}\epsilon^{\gamma\beta}\mathrm{d}x^{\dot\alpha\delta}\wedge\mathrm{d}x^{\dot\gamma\tau}\epsilon_{\delta\tau}+\tensor{R}{_\gamma^\alpha_{\dot\gamma}^{\dot\beta}}\epsilon^{\dot\alpha\dot\gamma}\mathrm{d}x^{\dot\delta\gamma}\wedge\mathrm{d}x^{\dot\tau\beta}\epsilon_{\dot\delta\dot\tau}.
\end{equation}
Hence, a given two form $\omega=\omega^\star_{\alpha\dot\alpha\beta\dot\beta}\star\mathrm{d}x^{\dot\alpha\star\alpha}\wedge_\star\mathrm{d}x^{\dot\beta\star\beta}$ can be split into its self-dual and anti-self dual parts $\omega_{SD}=\omega^\star_{\dot\alpha\dot\beta}\star(\mathrm{d}x^{\dot\alpha\alpha}\wedge\mathrm{d}x^{\dot\beta\beta}\epsilon_{\alpha\beta})$ and $\omega_{_{ASD}}=\omega^\star_{\alpha\beta}\star(\mathrm{d}x^{\dot\alpha\alpha}\wedge\mathrm{d}x^{\dot\beta\beta}\epsilon_{\dot\alpha\dot\beta})$, where
\begin{equation}\label{eq:ASD+SD}
    \begin{aligned}
        \omega^\star_{\dot\alpha\dot\beta}&=\omega^\star_{\alpha\dot\alpha\beta\dot\gamma}\tensor{R}{_\gamma^\alpha_{\dot\beta}^{\dot\gamma}}\epsilon^{\gamma\beta}&\qquad
        \omega^\star_{\alpha\beta}&=\omega^\star_{\gamma\dot\alpha\beta\dot\beta}\tensor{R}{_\alpha^\gamma_{\dot\gamma}^{\dot\beta}}\epsilon^{\dot\alpha\dot\gamma}.
    \end{aligned}
\end{equation}

Hence, this forms a basis of self-dual and anti-self-dual two-forms in the deformed setting. In the proceding sections, we will use this basis to explicitly express the self-dual and anti-self-dual sectors of noncommutative Yang-Mills theory in a given basis. This will be the four dimensional formulation, we will construct via the dimensional reduction from the appropriately deformed 6d BF theory, which will be constructed in section \ref{sec:DefBFTheory}.

\subsection{Noncommutative Yang Mills theory}\label{NonCommutativeYM}

Using the systematic differential calculus and Hodge dual found in \cite{Meier:twist} and reviewed in section \ref{sec:DiffCalc}, a consistent gauge invariant noncommutative Yang-Mills theory can be defined via introducing a gauge algebra valued one-form  $A$ which transforms under star gauge transformations as\footnote{Note that the star-gauge transformation do not close for arbitrary gauge algebras. Let us consider the star commutator of two gauge transformations $\epsilon_1$ and $\epsilon_2$
\begin{equation}
\scom{\epsilon_1}{\epsilon_2}=(\epsilon^a_1\star\epsilon^b_2- \epsilon^a_2\star\epsilon^b_1)T^aT^b,
\end{equation}
where $T^a$ denote a basis of the gauge algebra. The latter does not give rise to a pure commutator. As for example the trace of the right hand side is not vanishing even when $T^a\in\mathfrak{su}(n)$, the gauge transformations do not close for a gauge group $SU(n)$. Instead we would consider star-gauge transformations being generated by its universal enveloping algebra $U(\mathfrak{su}(n))$. By the Seyberg-Witten map the degrees of freedom in the universal enveloping algebra are mapped to the degrees of freedom of the undeformed gauge algebra \cite{Seiberg-Witten}.}
\begin{equation}
    \delta_\epsilon A=-\mathrm{d}\epsilon+\scom{\epsilon}{A}.
\end{equation}
Its field strength, which we denote by $G$ in order to avoid confusion with the Drinfel'd twist, is given by
\begin{equation}
    G=\mathrm{d}A+A\wedge_\star A,
\end{equation}
which transforms covariantly under star-gauge transformations. In components, the gauge field reads $A=A^\star_{\alpha\dot\alpha}\star \mathrm{d}x^{\dot\alpha\star\alpha}$, and therefore the fields strength is written as $G=G^\star_{\alpha\dot\alpha\beta\dot\beta}\star \mathrm{d}x^{\dot\alpha\star\alpha}\wedge_\star \mathrm{d}x^{\dot\beta\star\beta}$, where:
\begin{equation}\label{eq:gaugeTwoComponent}
G^\star_{\alpha\dot\alpha\beta\dot\beta} = \partial^\star_{\alpha\dot\alpha}A^\star_{\beta\dot\beta} + A^\star_{\gamma\dot\gamma}\star R_{\alpha}{}^{\gamma}R_{\dot\alpha}{}^{\dot\gamma}A^\star_{\beta\dot\beta}.
\end{equation}
Furthermore, due to star linearity of the Hodge star operator, the dual field strength is covariant as well, which leads to the gauge invariant Yang-Mills action \cite{Meier:twist}
\begin{equation}
    S_{\tiny \mbox{YM}}=\frac{1}{2g_{\tiny \mbox{YM}}^2}\int\tr{~G\wedge_\star*G}.
\end{equation}
Furthermore, we can add the noncommutative equivalent of the $\theta$-term:
\begin{equation}
    S_\theta=-\frac{\theta}{8\pi^2}\int\tr~G\wedge_\star G,
\end{equation}
which was argued to be topological as the integral of an exact top form \cite{Meier:twist}. By tuning $\theta$ appropriately, the action results in
\begin{equation}
    S_{YM\pm\theta}=\frac{-i}{4g_{\tiny \mbox{YM}}^2}\int\tr~(G\pm*G)\wedge_\star(G\pm*G),
\end{equation}
which is positive definite and hence solved by the noncommutative self-dual or anti-self-dual field equations respectively:
\begin{equation}
\label{eq:SDFieldEqs}
    G=\mp*G.
\end{equation}
By choosing the chiral basis for two-forms in \eqref{eq:SpinorBas2Form}, the field strength becomes
\begin{equation}
    G=G^\star_{\alpha\beta}\star(\mathrm{d}x^{\dot\alpha\alpha}\wedge\mathrm{d}x^{\dot\beta\beta}\epsilon_{\dot\alpha\dot\beta})+G^\star_{\dot\alpha\dot\beta}\star(\mathrm{d}x^{\dot\alpha\alpha}\wedge\mathrm{d}x^{\dot\beta\beta}\epsilon_{\alpha\beta}),
\end{equation}
where the ASD and SD components read respectively:
\begin{equation}
G^\star_{\dot\alpha\dot\beta}=G^\star_{\alpha\dot\alpha\beta\dot\gamma}\tensor{R}{_\gamma^\alpha_{\dot\beta}^{\dot\gamma}}\epsilon^{\gamma\beta},\ \ \ \ G^\star_{\alpha\beta}=G^\star_{\gamma\dot\alpha\beta\dot\beta}\tensor{R}{_\alpha^\gamma_{\dot\gamma}^{\dot\beta}}\epsilon^{\dot\alpha\dot\gamma},
\end{equation}
and \eqref{eq:SDFieldEqs} becomes
\begin{equation}
\begin{aligned}
    G^\star_{\dot\alpha\dot\beta}&=0\\
    G^\star_{\alpha\beta}&=0,
\end{aligned}
\end{equation}
respectively. The anti-self-dual field strength is then given by:
\begin{equation}\label{ASDG*}
G_{_{ASD}} = G^\star_{\alpha\beta}\star(\mathrm{d}x^{\dot\alpha\alpha}\wedge \mathrm{d}x^{\dot\beta\beta}\epsilon_{\dot\alpha\dot\beta}).
\end{equation}
By considering the anti-self dual action, we can introduce an auxiliary field $B_{}=B_{\alpha\beta}^\star\star(\mathrm{d}x^{\dot\alpha\alpha} \wedge\mathrm{d}x^{\dot\beta\beta}\epsilon_{\dot\alpha\dot\beta})$, such that
\begin{equation}\label{noncomutativeaction}
    S^\star_{YM-\theta}=\int\tr~B_{}\wedge_\star G_{_{ASD}} + \frac{g^2_{YM}}{4} B_{}\wedge_\star B_{},
\end{equation}
which will be the action we are going to reproduce from a reduction of the BF theory on $\mathbb{CP}^3$ to 4d Minkowski space in the next section.

\subsection{Matter Fields}\label{sec:MatterFields}

Now that we have reviewed the noncommutative Yang-Mills theory, we will review how it can be coupled to matter fields. We will consider scalar and fermionic fields in the adjoint representation, as it is what appears in the ${\cal N}=4$ SYM theory. Importantly, we will formulate the action using an index free notation, since manifestly transform properly under star gauge symmetry after the deformation. 
\paragraph{Scalar fields}
A scalar field in the adjoint representation transforms as
\begin{equation}
\delta_\epsilon\phi = [\epsilon \stackrel{\star}{,} \phi].
\end{equation}
The covariant derivative is correspondingly given by:
\begin{equation}
D\phi = d\phi +[A\stackrel{\star}{,} \phi]
\end{equation}
Then, the noncommutative kinetic action for the scalar fields is:
\begin{equation}
S^*_{\phi} = \int \tr \Big(D\phi^\dagger\wedge_\star *D\phi\Big),
\end{equation}
The interaction terms involving scalar fields can be added by replacing products by star products of the usual expressions. For instance, the $\phi^4$ interaction reads:
\begin{equation}
\int \mathrm{d}^4x\ \tr\Big(\phi\star\phi^\dagger\star\phi\star\phi^\dagger\Big).
\end{equation}

\paragraph{Fermionic fields}
In order to define gauge invariant actions including fermionic fields, we need to introduce grassmann-odd basis spinors $s^\alpha$ and $\bar s^{\dot\alpha}$. Similar to the gauge field $A_\mu$, Weyl spinors carry a nontrivial representation under the Poincar\'e algebra. Similar to the basis one-forms, the basis spinors will combine with the component Weyl spinors to index-free objects, which will be considered to be the fundamental objects transforming under star-gauge symmetry \cite{Meier:twist}. For this purpose, a given vector field will act on the basis spinors by the spinorial Lie derivative \cite{SpinorLie} as
\begin{align}
\begin{split}
&{\cal L}_X(s^\alpha) = -\frac{1}{8} (\partial^{\alpha\dot\alpha} X_{\beta\dot\alpha} - \partial_{\beta\dot\alpha} X^{\alpha\dot\alpha})s^\beta\\
&{\cal L}_X(\bar s^{\dot\alpha}) = -\frac{1}{8}(\partial^{\alpha\dot\alpha} X_{\alpha\dot\beta} - \partial_{\alpha\dot\beta} X^{\alpha\dot\alpha})\bar s^{\dot\beta}
\end{split}
\end{align}
We then introduce index free spinors as
\begin{align}\label{eq:HalfFormsPsiDeformed}
\begin{split}
&\psi = s^\alpha \psi_\alpha = s^\alpha\star\psi^\star_\alpha,\\
&\bar\psi = \bar\psi_{\dot\alpha}\bar s^{\dot\alpha} = \bar\psi^\star_{\dot\alpha}\star\bar s^{\dot\alpha},
\end{split}
\end{align}
where $\psi^\star_\alpha$ and $\bar\psi^\star_{\dot\alpha}$ are the Grassmann-odd left-handed and right-handed component spinors respectively. The integrating over the basis spinors results in
\begin{equation}\label{eq:IntSpinorBasis}
\int d^2s\  s^\alpha s^\beta = \epsilon^{\alpha\beta},\ \ \ \ \ \int d^2\bar s\ \bar s^{\dot\alpha}\bar s^{\dot\beta} = -\epsilon^{\dot\alpha\dot\beta}.
\end{equation}
The algebra of basis spinors is similar to the star-commutation rules of differential forms:
\begin{equation}
\label{eq:Fwithspinorindices}
\begin{aligned}
f(x) \star s^\alpha&=s^\beta \left(\bar{F}_{op}\right)_\beta^{\enspace\alpha} f(x) &&&
f(x) \star \bar{s}^{\dot{\alpha}}&=\bar{s}^{\dot{\beta}} \left(\bar{F}_{op}\right)_{\dot{\beta}}^{\enspace\dot{\alpha}} f(x)\\
\bar{s}^{\dot{\alpha}}\star s^\alpha&=\bar{F}^{\enspace\dot{\alpha}\enspace \alpha}_{\dot{\beta}\enspace\beta} \bar{s}^{\dot{\beta}} s^\beta &&&
s^\alpha \star \bar{s}^{\dot{\alpha}}&=\bar{F}^{\enspace \alpha \enspace \dot{\alpha}}_{\beta \enspace \dot{\beta}} s^\beta \bar{s}^{\dot{\beta}}.\\
s^{\alpha}\star s^\beta&=\bar{F}^{\enspace\alpha\enspace \beta}_{\gamma\enspace\delta} \bar{s}^{\gamma} s^\delta &&&
\bar{s}^{\dot{\alpha}} \star \bar{s}^{\dot{\beta}}&=\bar{F}^{\enspace \alpha \enspace \beta}_{\gamma \enspace \delta} \bar{s}^{\dot{\gamma}} \bar{s}^{\dot{\delta}},
\end{aligned}
\end{equation}
We can now also move functions and basis one-forms through the basis half forms, which introduces $\mathcal{R}$ matrices with appropriate spinor indices as
\begin{equation}\label{eq:RForSpinorBasis}
\begin{aligned}
s^\alpha \star f(x)&=R_\beta{}^\alpha f(x)\star s^\beta&&&
\bar{s}^{\dot{\alpha}} \star f(x)&=R_{\dot{\beta}}{}^{\dot{\alpha}} f(x)\star \bar{s}^{\dot{\beta}},\\
\bar{s}^{\dot{\alpha}}\star s^\alpha&=-R^{\enspace\dot{\alpha}\enspace \alpha}_{\dot{\beta}\enspace\beta} s^\beta \star \bar{s}^{\dot{\beta}}&&&
s^\alpha \star \bar{s}^{\dot{\alpha}}&=-R^{\enspace \alpha \enspace \dot{\alpha}}_{\beta \enspace \dot{\beta}} \bar{s}^{\dot{\beta}} \star s^\beta,\\
s^\alpha\star s^\beta&=-R^{\enspace\alpha\enspace \beta}_{\gamma\enspace\delta} s^\delta \star s^\gamma&&&
\bar{s}^{\dot{\alpha}} \star \bar{s}^{\dot{\beta}}&=-R^{\enspace \dot{\alpha} \enspace \dot{\beta}}_{\dot{\gamma} \enspace \dot{\delta}} \bar{s}^{\dot{\delta}} \star \bar{s}^{\dot{\gamma}}.
\end{aligned}
\end{equation}
Note that the twists and $R$-matrices appearing in \eqref{eq:Fwithspinorindices} and \eqref{eq:RForSpinorBasis} are identical to the ones appearing in \eqref{eq:HandedRMatRep}. This defines the noncommutative structure upon which we want to define our gauge theory. Similar to the gauge transformation for the gauge field, we will treat the full spinors $\psi$ and $\bar{\psi}$ as the fundamental objects to assign gauge transformations.

Furthermore, we introduce the one-form  $\sigma$ defined as the index-free object corresponding to the Pauli-matrices as
\begin{equation}\label{eq:OneFormSigmaDeformed}
\sigma = s^\alpha \bar s^{\dot\alpha}\mathrm{d}x_{\alpha\dot\alpha} = s^\alpha\star\bar s^{\dot\alpha}\star \mathrm{d}x_{\dot\alpha\star\alpha},
\end{equation}
where the deformed basis one-form  $\mathrm{d}x_{\dot\alpha\star\alpha}$ is defined by equation (\ref{eq:SpinorBas2Form}). This one-form  exhibits the crucial property of star-commutativity, which is essential for the construction of a gauge-invariant noncommutative action for the fermionic field $\psi$:
\begin{equation}\label{eq:SigmaCommutes}
\sigma \star f = f \star \sigma.
\end{equation}

With the algebraic structure of the half-forms in place, we can now formulate a gauge theory with fermionic matter by treating $\psi$ and $\bar \psi$ as the fundamental objects under gauge transformations. We consider the adjoint fermionic fields, which transforms as:
\begin{equation}
\delta_\epsilon\psi = [\epsilon \stackrel{\star}{,} \psi],
\end{equation}
which leads to the covariant derivative:
\begin{equation}
D\psi = d\psi -i[A\stackrel{\star}{,} \psi].
\end{equation}
Then, the noncommutative kinetic action for the fermionic fields is:
\begin{equation}
S^\star_{\psi} = \int d^2s d^2\bar s \int \tr \Big( \bar\psi \star\sigma \wedge_\star *D\psi\Big),
\end{equation}
Again, we can add interaction terms involving scalar and fermionic fields by replacing products by star products
of the usual expressions. For instance, the Yukawa-like interactions reads as:
\begin{equation}
\int d^2s\int \mathrm{d}^4x\ \tr\Big( \psi\star\phi\star\psi\Big).
\end{equation}

\subsection{Noncommutative ${\cal N}=4$ SYM}\label{NCSYM}

The construction developed in this chapter enables a noncommutative formulation of ${\cal N}=4$ super-Yang-Mills theory. Its field content consists of a gauge field $A$; four fermionic fields and their antifields $\Psi^i$ and $\bar\Psi_i$; and six scalar fields $\Phi^{ij}$, antisymmetric in the indices $i,j=1,\cdots,4$. The action is given by:
\begin{align}\label{eq:ActionDefSYM}
\begin{split}
S^\star_{\tiny \mbox{SYM}-\theta} &=\int  \tr\Big(B_{}\wedge_\star G_{_{ASD}} + \frac{g^2_{\tiny\mbox{YM}}}{4}B_{}\wedge_\star B\Big)\\
&\ \ \ + \int \tr\Big(D\Phi^{ij}\wedge_\star *D\Phi_{ij}\Big) + \int d^2s d^2\bar s\ \tr\Big(\bar\Psi_i \star \sigma \wedge_\star *D\Psi^i \Big)\\
&\ \ \ + \int d^2\bar s \int \mathrm{d}^4x\ \tr\Big( \bar\Psi_i \star [\Phi^{ij}\stackrel{\star}{,}\bar\Psi_j] \Big) + \int d^2s\ \int \mathrm{d}^4x\ \tr\Big(\Psi^i\star[\Phi_{ij}\stackrel{\star}{,}\Psi^j]\Big)\\
&\ \ \ +  \int \mathrm{d}^4x\ \tr\Big( [\Phi^{ij}\stackrel{\star}{,}\Phi^{kl}]\star[\Phi_{ij}\stackrel{\star}{,}\Phi_{kl}]\Big)
\end{split}
\end{align}

\section{Twisting twistor space}\label{sec:DefTwistor}

In this section, we will define a twisted twistor space along the lines of the construction of twist deformed spacetime in the previous section. We will then define a noncommutative version of BF theory, which is invariant under star gauge symmetry and will reduce to the noncommutative anti-self-dual Yang-Mills theory presented in the latter section. Furthermore, we will construct a twisted supersymmetric extension, which is perturbatively equivalent to $\mathcal{N}=4$ SYM.

\paragraph{The Twist in twistor space} In twistor space, the conformal symmetry of the four dimensional space manifests as an $SL(4)$ symmetry of the twistor space. By dividing the four homogeneous coordinates $Z^i$ into a right-handed and a left-handed spinor $\mu_{\dot\alpha}$ and $\lambda^\alpha$, the $SL(4)$ symmetry splits naturally into Poincar\'e, scale and special conformal transformations after imposing the incidence relation \eqref{eq:incidence}. In this way, for every twist in the conformal algebra, there is a corresponding twist of the $SL(4)$ symmetry of the twistor space dictated by the incidence relation that is imposed to relate the twistor space to four-dimensional spacetime.

The full conformal algebra acts linear on the homogeneous coordinates of twistor space, however, imposing the incidence relation makes the Poincar\'e subalgebra special. In fact, it is leaving the residual $\mathbb{CP}^1$ invariant, while special conformal and scale transformations mix the four spacetime coordinates with the coordinates on the remaining $\mathbb{CP}^1$. Although there is no general caveat with constructing a gauge invariant BF theory on twisted $\mathbb{CP}^3$ for a twist of the full $SL(4)$ symmetry with our construction, we will restrict to twists of the Poincar\'e algebra. The reason for this is that the intact $\mathbb{CP}^1$ residual space is desirable for integrating out the fibre dependence and hence allows a reduction of the 6 dimensional theory to four dimensional gauge theory. Moreover, the (anti-)self-dual basis of differential forms in the undeformed theory still form an (anti-)self-dual basis in the deformed theory for twists in the Poincar\'e algebra as shown in \eqref{eq:AntiSelfDualUndeformed}. For twists involving special conformal and dilatation generators this might not be the case anymore.

\subsection{Noncommutative twistor space}\label{subsec:NCTwistor}
In the setup of the four dimensional theory, the Drinfel'd twist is considered to act via Lie derivatives on fields, functions and differential forms. Similarly, we will let the twist act as Lie derivatives on fields and differential forms on twistor space. In order to find an algebra of differential forms on twistor space, that is a Hopf algebra module of the twisted $\mathfrak{sl}(4)$ symmetry, we need to replace products by star products again. This leads to a noncommutative twistor space where the coordinates satisfy the commutation relations
\begin{equation}
\begin{aligned}
 Z^i\star Z^j=\tensor{R}{_k^i_l^j}Z^l\star Z^k.
\end{aligned}
\end{equation}
Note that we keep the incidence relation and hence the way to identify four dimensional space with twistor space undeformed. The reason for this choice is that the Drinfel'd twist naturally deforms products of objects, while the representation under the symmetries of a single vector space remains undeformed. The coordinates $Z^i$ are fundamental w.r.t. our twist and hence $\mu^{\dot\alpha}$ on its own remain the same in the deformed theory w.r.t. the action of the $SL(4)$ symmetry. Furthermore,  in the spacetime theory there is always the ambiguity to express a differential form w.r.t. a standard basis of differential forms without any star product or a basis of star-product forms. We treat the incidence relation analogously.

The main reason for restricting to deformations of the Poincar\'e algebra only, is that we can observe some simplifications in the latter star product after imposing the incidence relation. In the Poincar\'e algebra, only the left $\mathfrak{sl}(2)$ acts non-trivially on the left-handed spinors $\lambda^\alpha$, particularly translations do not affect left-handed spinors. A twist in the Poincar\'e algebra that is twist unimodular and hence gives rise to a cyclic star product is not allowed to contain any term that is acting with both its legs nontrivially on pure left-handed spinors. More explicitly, writing the twist in Sweedler notation $\mathcal{F}=f_\alpha\otimes f^\alpha$, for each $\alpha$ if $f_\alpha\in U(\mathfrak{sl}(2)_L)$ then $f^\alpha\in U(\mathfrak{g})$, where $\mathfrak{g}$ is orthogonal to the $\mathfrak{sl}(2)_L$ similarly with $f_\alpha$ and $f^\alpha$ being exchanged. This can be verified explicitly given the classification of classical r-Matrices in the Poincar\'e algebra in \cite{Zakrzewski:1997mna}, cf. \cite{Meier:twist} for all examples leading to cyclic star products. Hence, in every such twist, there must be always one leg, that acts trivially on left-handed spinors. Thus, the commutation relation of two left-handed spinors simplifies as
\begin{equation}\label{l*l}
    \lambda^\alpha\star\lambda^\beta=\lambda^\beta\star\lambda^\alpha=\lambda^\alpha\lambda^\beta.
\end{equation}
This directly means that the residual $\mathbb{CP}^1$ remains undeformed and commutative. As a result, the integration over the fibre in the deformed setting will be similar as in the undeformed setting. Despite the commutativity of the $\mathbb{CP}^1$ coordinates among each other, they do not commute with the spacetime coordinates $x^{\alpha\dot\alpha}$. Hence, a left-handed spinor commutes with any object up to an R matrix, which takes the form of an operator valued left-handed Lorentz transformation matrix:
\begin{equation}
    \lambda^\alpha\star f=\tensor{R}{_\beta^\alpha}(f)\star \lambda^\beta.
\end{equation}
Note that $\tensor{R}{_\beta^\alpha}$ coincides with the left-handed part of the R-matrix representation defined for the spinor index basis in the four dimensional spacetime \eqref{eq:HandedRMatRep}.

It is desirable to have a similar behavior for the right-handed spinor. However, as those do not only transform under the right $\mathfrak{sl}(2)$ but also under translations, commuting a right-handed spinor over a function is not captured by the right-handed R matrix $\tensor{R}{_{\dot\alpha}^{\dot\beta}}$ only. We will comment on how to recast the right commutation law for the desired objects after defining a basis of differential forms on twisted twistor space.

\paragraph{The incidence relation} Like in the undeformed twistor space, the incidence relation will define a map between $\mathbb{CP}^1$ curves in twistor space and spacetime positions. In fact, we will keep the identification intact even after turning on the deformation. Hence, the incidence relation is still given by
\begin{equation}
    \mu^{\dot\alpha}=x^{\dot\alpha\alpha}\lambda_\alpha,
\end{equation}
however, it will be usefull to rewrite this relation in terms of a star product in order to keep the twisted module structure under the twisted Hopf algebra. Hence, we define a twisted coordinate $\chi^{\dot\alpha\star\alpha}$, such that
\begin{equation}
    \label{eq:defIncidence}\mu^{\dot\alpha}=x^{\dot\alpha\alpha}\lambda_\alpha=\chi^{\dot\alpha\alpha}\star\lambda_\alpha
\end{equation}
and hence
\begin{equation}
\label{eq:defCP1Lines}
    \chi^{\dot\alpha\alpha}=\frac{\hat\mu^{\dot\alpha}\star\lambda^\alpha-\mu^{\dot\alpha}\star\hat\lambda^\alpha}{\braket{\lambda\stackrel{\star}{,}\hat\lambda}}.
\end{equation}
Note that $\chi^{\dot\alpha\alpha}\neq x^{\dot\alpha\star\alpha}$ in general. The star product apearing between indices is giving a twist with four indices, which is defined by acting with a twist on basis forms or basis spinors. Translation generators, however, act trivially on such objects and hence do not contribute to the star product between indices. In the star products in \eqref{eq:defIncidence} and \eqref{eq:defCP1Lines} between $\lambda^\alpha$ and $\mu^{\dot\alpha}$ and $\chi^{\dot\alpha\alpha}$, respectively, translation generators appearing in the twist can instead act on $\mu^{\dot \alpha}$ and $\chi^{\dot\alpha\alpha}$ and hence, the star product does not only reduce to the previously defined twists with four indices and therefore cannot be compensated by only inserting a star product between the indices as in $x^{\dot\alpha\star\alpha}$.

\paragraph{Basis of anti-holomorphic differential forms and vectors} Since the deformation process only affects the products of coordinates but not the parametrization of spacetime, also the vector spaces of one-form  and vector fields do not change. Hence, we can use the same basis as in the undeformed case, however, similarly, it is desirable to reformulate them manifestly using star products. For the anti-holomorphic part of the $\mathbb{CP}^1$ part, we choose the basis
\begin{equation}
\begin{aligned}
    \bar\partial_0&=\braket{\lambda,\hat\lambda}\lambda^\alpha\frac{\partial}{\partial\hat\lambda^\alpha}&=&\braket{\lambda\stackrel{\star}{,}\hat\lambda}\star\lambda^\alpha\star\frac{\partial}{\partial\hat\lambda^\alpha}\\
    \bar e^0&=\frac{\braket{\hat\lambda,\mathrm{d}\hat\lambda}}{\braket{\lambda,\hat\lambda}^2}&=&\frac{\braket{\hat\lambda\stackrel{\star}{,}\mathrm{d}\hat\lambda}}{\braket{\lambda\stackrel{\star}{,}\hat\lambda}^2},
\end{aligned}    
\end{equation}
where we used that left-handed spinors star commute with each other. Moreover, both are star commutative, as they are invariant under Poincar\'e transformations.\footnote{In fact, by restricting the class of deformations to the Poincar\'e algebra, we ensure first, that the residual $\mathbb{CP}^1$ stays undeformed and second, that the measure on the $\mathbb{CP}^1$ space is star-commuting. This will allow to recast the four dimensional gauge theory on twisted noncommutative spacetime later on.}

For the remaining part of the anti-holomorphic basis of vector fields and one-form , we choose the basis
\begin{equation}
\label{eq:partialDerivs}
    \begin{aligned}
        \bar\partial_{\dot\alpha}&=\lambda^\alpha\frac{\partial}{\partial x^{\alpha\dot\alpha}}\\
        \bar e^{\dot\alpha}&=\mathrm{d}x^{\dot\alpha\alpha}\hat\lambda_\alpha.
    \end{aligned}
\end{equation}
Again, we would like to insert star products to fit to the twisted structures. For this, first note that both $\frac{\partial}{\partial x^{\alpha\dot\alpha}}$ and $\mathrm{d}x^{\dot\alpha\alpha}$ are translation invariant. Hence, by inserting a star product between the latter and $\lambda^\alpha$, only the right-handed Lorentz generators in the twist can act nontrivially. Hence,
\begin{equation}\label{eq:partialDerivsStar}
\begin{aligned}
    \bar\partial_{\dot\alpha}&=\lambda^\beta\star\tensor{F}{_\beta^\alpha}\partial_{\alpha\dot\beta}\\
    \bar e^{\dot\alpha}&=\tensor{F}{_{\dot\beta}^{\dot\alpha}_\alpha^\beta}\mathrm{d}x^{\dot\beta\alpha}\star\hat\lambda_\beta=\mathrm{d}x^{\dot\alpha\star\alpha}\star\hat\lambda_\alpha,
\end{aligned}
\end{equation}
where this time $\mathrm{d}x^{\dot\alpha\star\alpha}$ is exactly as defined in \eqref{eq:SpinorBas2Form}. Note that only the right $\mathfrak{sl}(2)$ copy of the Lorentz algebra acts nontrivially on $\bar e^{\dot\alpha}$. Hence, we can commute a function over the basis form $\bar e^{\dot\alpha}$ up to an R-matrix as follows
\begin{equation}
    \bar e^{\dot\alpha}\star f=\tensor{R}{_{\dot\beta}^{\dot\alpha}}f\star \bar e^{\dot\alpha}.
\end{equation}

\paragraph{Exterior algebra} Similarly to the exterior algerbra on four-dimensional spacetime, we introduce a star-wedge product on twistor space, which again is totally R-antisymmetric. As $\bar e^0$ is Poincar\'e invariant, it automatically anticommutes with $\bar e^{\dot\alpha}$. Furthermore, as $\bar e^{\dot\alpha}$ only transforms under one copy of the $\mathfrak{sl}(2)$ of the Lorentz symmetry, a twist acting on two such one-form  will always act trivially and hence
\begin{equation}\label{e^star^e}
    \bar e^{\dot\alpha}\wedge_\star\bar e^{\dot\beta}=-\bar e^{\dot\beta}\wedge_\star\bar e^{\dot\alpha}=\bar e^{\dot\alpha}\wedge\bar e^{\dot\beta}=\frac{1}{2}\epsilon^{\dot\alpha\dot\beta}\bar e^{\dot\gamma}\wedge\bar e_{\dot\gamma}=\frac{1}{2}\epsilon^{\dot\alpha\dot\beta}\bar e^{\dot\gamma}\wedge_\star\bar e_{\dot\gamma}.
\end{equation}
Hence, the anti-holomorphic top form remains undeformed:
\begin{equation}\label{Omegastar}
    \frac{\bar\Omega^3}{\braket{\lambda\hat\lambda}^4}=\bar e^0\wedge_\star\bar e^{\dot\alpha}\wedge_\star\bar e_{\dot\alpha}=\bar e^0\wedge\bar e^{\dot\alpha}\wedge\bar e_{\dot\alpha}.
\end{equation}
\paragraph{Complex structure} Before we start discussing gauge theories on twisted twistor space, we need to define a complex structure and with it an anti-holomorphic differential operator. Similarly to the discussion of the exterior derivative in the context of the four dimensional deformed spacetime, the undeformed anti-holomorphic differential operator $\bar\partial=\bar e^0 \bar\partial_0+\bar e^{\dot\alpha}\bar\partial_{\dot\alpha}$ remains an integrable almost complex structure. Furthermore, it commutes with the Lie derivatives w.r.t. the $\mathfrak{sl}(4)$ symmetry generators and hence is not affected by the twist. Therefore, it satisfies the normal Leibniz rule when acting on star products of forms:
\begin{equation}
    \bar\partial(\omega\wedge_\star\chi)=\bar\partial\omega\wedge_\star\chi+(-1)^{\abs{\chi}}\omega\wedge_\star\bar\partial\chi.
\end{equation}

Similarly to the discussion of the exterior derivative in the 4d spacetime, we can express the anti-holomorphic derivative w.r.t. a star product basis or a normal linear basis and hence
\begin{equation}
    \bar\partial f=\bar\partial^\star_0f\star\bar e^0+\bar\partial_{\dot\alpha}^\star f\star\bar e^{\dot\alpha}=\bar\partial_0f\bar e^0+\bar\partial_{\dot\alpha} f\bar e^{\dot\alpha},
\end{equation}
where
\begin{equation}
    \begin{aligned}
        \bar\partial_0^\star f&=\bar\partial_0 f\\
        \bar\partial_{\dot\alpha}^\star f&=\tensor{F}{_{\dot\alpha}^{\dot\beta}}\bar\partial_{\dot\beta}f.
    \end{aligned}
\end{equation}
We can relate the star derivative in the twistor space to the star derivative in four dimensions by inserting the equation \eqref{eq:partialDerivsStar}:
\begin{equation}
    \begin{aligned}
        \bar\partial_{\dot\alpha}^\star f&=\tensor{\left.F_{op}\right.}{_{\dot\alpha}^{\dot\beta}}(\tensor{\left.F_{op}\right.}{_\beta^\alpha}\partial_{\alpha\dot\beta}f\star\lambda^\beta)\\
        &=\tensor{F}{_{\alpha'}^\alpha_{\dot\beta'}^{\dot\beta}}\tensor{\left.F_{op}\right.}{_{\dot\alpha}^{\dot\beta'}}\tensor{\left.F_{op}\right.}{_\beta^{\alpha'}}\partial_{\alpha\dot\beta}f\star\lambda^\beta\\
        &=\partial^\star_{\alpha\star\dot\alpha}f\star\lambda^\alpha.
    \end{aligned}
\end{equation}

\subsection{Gauge theory on noncommutative twistor-space}

The twistor space reviewed in section \ref{sec:TwistorSpace} provides a particularly convenient framework for describing the anti–self-dual sector of Yang–Mills theory. Indeed, because of its index structure, this formulation allows an explicit separation between the self-dual (SD) and anti–self-dual (ASD) sectors. As discussed in Section \ref{sec:DefBFTheory}, it turns out that even in the twisted noncommutative space it is possible to construct a purely chiral basis, consisting of pure left-handed and right-handed two-forms, as shown in equation \eqref{eq:PureChiralBasisDeformed}.

In this section, we begin with a twisted gauge theory formulated on the 6-dimensional twisted twistor space, and then we show that it is equivalent to the four-dimensional noncommutative Yang–Mills theory, perturbatively around the ASD sector. Similar to the construction of noncommutative Yang-Mills theory in spacetime as described in section \ref{sec:DefBFTheory}, we will deform the twistor theory by replacing products by star products in the appropriate description in order to obtain a star gauge invariant action. We will introduce the star-gauge equivalents of the gauge one-form $a$ and the auxiliary one-form $b$. Moreover, we will construct both, the local deformed action $S_1^\star[a,b]$ as well as the nonlocal action $S_2^\star[b]$. We will show its invariance under star gauge symmetries and in the second part of this section, we will reduce the twistor gauge theory to the noncommutative gauge theory in four dimensional spacetime.

\paragraph{Gauge invariant action}
The deformation of the local action encoding the ASD YM equation is given as
\begin{equation}\label{BF*}
S^\star_1[a,b] = \frac{i}{2\pi}\int \Omega^3\wedge_\star \tr\left[ b\wedge_\star\left(\bar\partial a + a\wedge_\star a\right) \right],
\end{equation}
where the individual fields transform under star-gauge transformations as
\begin{equation}
\delta_\epsilon a = \bar\partial\epsilon + [\epsilon\stackrel{\star}{,}a],\ \ \ \delta_\epsilon b = [\epsilon\stackrel{\star}{,}b],
\end{equation}
for $\epsilon$ an element of the gauge algebra. As a result, the local term $S_1$ transforms as
\begin{equation}\label{eq:starGaugetfS1}
\delta_\epsilon S^\star_1[a,b] = \frac{i}{2\pi}\int \tr~\Omega^3\wedge_\star \left[\epsilon\stackrel{\star}{,}b\wedge_\star\left(\bar\partial a + a\wedge_\star a\right)\right].
\end{equation}
Note that the top form $\Omega^3$ is invariant under the $SL(4)$ symmetry and hence it is star-commutative, i.e.
\begin{equation}
    \Omega^3\star f=f\star\ \Omega^3
\end{equation}
for every function $f$. Hence, equation \eqref{eq:starGaugetfS1} becomes
\begin{equation}
\delta_\epsilon S^\star_1[a,b] = \frac{i}{2\pi}\int\tr \left[\epsilon\stackrel{\star}{,}\Omega^3\wedge_\star b\wedge_\star\left(\bar\partial a + a\wedge_\star a\right)\right].
\end{equation}
The latter is the integral and trace over a star commutator and hence vanishes if the star product is cyclic under integration, similar to the Yang-Mills action in four dimensions, cf. section \ref{NonCommutativeYM}.

Analogously, we can define a noncommutative counterpart to the non-local action as
\begin{equation}
\begin{aligned}\label{eq:defS2}
S_2^\star[b] &= \frac{g^2_{\tiny \mbox{YM}}}{8\pi^2}\int\mathrm{d}^4x\int_{\mathbb{CP}^1\times \mathbb{CP}^1}\omega_1\wedge_\star\omega_2\ \star\langle\lambda_1\lambda_2\rangle^2\star  \tr(b_1\star b_2)\\
&=\frac{g^2_{\tiny \mbox{YM}}}{8\pi^2}\int\mathrm{d}^4x\int_{\mathbb{CP}^1}\omega_1\int_{\mathbb{CP}^1}\omega_2\ \langle\lambda_1\lambda_2\rangle^2~  \tr(b_1\star b_2).
\end{aligned}
\end{equation}
Note that $\omega_i$ and $\langle\lambda_1\lambda_2\rangle$ are Lorentz scalars and hence star commutative for twists in the Poincar\'e algebra. Its gauge-invariance can be similarly verified. Explicitly, it is given by
\begin{equation}
\delta_\epsilon S_2^\star[b] = \frac{g^2_{\tiny \mbox{YM}}}{8\pi^2}\int \mathrm{d}^4x\int_{\mathbb{CP}^1}\omega_1\int_{\mathbb{CP}^1}\omega_2\ \langle\lambda_1\lambda_2\rangle^2\ \tr\Big(\left[\epsilon\stackrel{\star}{,}b_1\star b_2\right]\Big) = 0,
\end{equation}
where we also used the $\star$-cyclicity under integration in the last equality above and that $\Braket{\lambda_1\lambda_2}$ is star-commutative. The full twistor gauge-invariant action is then given by the sum of the two terms:
\begin{equation}\label{eq:FullTwistorAction}
S^\star[a,b] = S_1^\star[a,b] + S_2^\star[b]
\end{equation}

\paragraph{Integrating the fibre}

Having constructed the gauge-invariant action on a noncommutative twistor space. We will now show that, after integrating over the $\mathbb{CP}^1$ fibers, this action is equivalent to the noncommutative Yang Mills theory (\ref{noncomutativeaction}). The procedure is motivated from the undeformed case reviewed in section \ref{subsec:twistorYM}. We first fix a gauge in order to be able to make the dependence of the individual fields on the fiber directions manifest. Applying the ideas of noncommutative gauge theories to the twistor space action, we first express the fields $a$ and $b$ using the twisted basis:
\begin{equation}\label{eq:YMFieldUsingStarProduct}
a = a_0^\star\star\bar e^0 + a_{\dot\alpha}^\star\star\bar e^{\dot\alpha},\ \ \ \ b = b_0^\star\star\bar e^0 + b_{\dot\alpha}^\star\star\bar e^{\dot\alpha}
\end{equation}
We now choose a gauge similar to the undeformed case by imposing\footnote{This choice of a gauge remains possible for two reasons. First $\bar e^0$ and $e^0$ are star-commutative and hence, the zero component of the fields and derivatives never mix with others. Second, if we start with a field configuration $a^{\prime 0}$, that does not satisfy \eqref{eq:gaugeCondStar}, there is always a gauge transformation, that satisfies
\begin{equation}
    \bar\partial^\star|_X\bar D_0\epsilon=\bar\partial^\star_0 D_0\epsilon=-\bar\partial^\star_0 a^{\prime 0}=-\bar\partial^\star|_Xa^{\prime 0},
\end{equation}
such that $a^0=a^{\prime 0}+D_0\epsilon$ satisfies the gauge condition \eqref{eq:gaugeCondStar} like in the undeformed case.}
\begin{equation}\label{eq:gaugeCondStar}
\bar\partial^*\big|_X a^\star_0=\bar\partial^*\big|_X b^\star_0=0,
\end{equation}
for $X\cong\mathbb{CP}^1$, where $\bar\partial^*|_X$ is the adjoint of the operator $\bar\partial|_X = \bar e^0\bar\partial_0$, as used previously in (\ref{eq:gaugeconditionYM}) for the undeformed case. Again, this gauge condition implies that $a_0^\star$ and $b_0^\star$ are harmonic, so the gauge fixing condition forces the fields to take the form:
\begin{align}\label{A0=0}
&a_0^\star = 0 ,\ \ \ \ \ \ b_0^\star = 3B_{\alpha\beta}^\star\star \frac{\hat\lambda^\alpha\star\hat\lambda^\beta}{\langle\lambda\hat\lambda\rangle^2}.
\end{align}
Hence, $a$ and $b$ become
\begin{align}\label{eq:GaugeFixingFieldsBFDef}
a = a^\star_{\dot\alpha}(x,\lambda,\hat\lambda)\star\bar e^{\dot\alpha},\ \ \ \ b = 3B_{\alpha\beta}^\star\star \frac{\hat\lambda^\alpha\star\hat\lambda^\beta}{\langle\lambda\hat\lambda\rangle^2}\star\bar e^0 +b^\star_{\dot\alpha}(x,\lambda,\hat\lambda)\star\bar e^{\dot\alpha}
\end{align}
after gauge fixing. Plugging these expressions into the action (\ref{BF*}), we get:
\begin{align}\label{1stepS}
\begin{split}
S_1^\star[a,B] = &\frac{i}{2\pi}\int \Omega^3\wedge_\star\tr\left[ 3B_{\alpha\beta}^\star\star \frac{\hat\lambda^\alpha\star\hat\lambda^\beta}{\langle\lambda\hat\lambda\rangle^2}\star\bar e^{0}\wedge_\star\left(\bar\partial^\star_{\dot\gamma} a^\star_{\dot\delta}\star\bar e^{\dot\gamma}\wedge_\star\bar e^{\dot\delta} + a^\star_{\dot\gamma}\star\bar e^{\dot\gamma}\wedge_\star a^\star_{\dot\delta}\star\bar e^{\dot\delta}\right)  \right]\\
&+ \frac{i}{2\pi}\int \Omega^3\wedge_\star\tr\Big[ b^\star_{\dot\alpha}\star\bar e^{\dot\alpha}\wedge_\star\bar\partial^\star_0 a^\star_{\dot\beta}\star\bar e^0\wedge_\star\bar e^{\dot\beta}  \Big],
\end{split}
\end{align}
The field $B^\star_{\dot\alpha}$ appears only linearly in the action above, and therefore it plays the role of a Lagrange multiplier again. Integrating out results in
\begin{equation}
\bar\partial_0^\star a^\star_{\dot\beta} = \bar\partial_0 a^\star_{\dot\beta} = 0 \Rightarrow a^\star_{\dot\alpha}(x,\lambda,\hat\lambda) = a^\star_{\dot\alpha}(x,\lambda).
\end{equation}
As in the undeformed case, since $a$ is of homogeneous degree 0 and $\bar e^{\dot\alpha}$ has degree $+1$, it follows that\footnote{We can express the expansion of $a$ in powers of $\lambda$ with ordinary products as well, such that $a^\star_{\dot\alpha}=A_{\alpha\dot\alpha}\lambda^\alpha $, however it will be more convenient to directly work in star products than in ordinary products. Both conventions are related to each other by $\tensor{\left.\bar F_{op}\right.}{_\alpha^\beta}A^\star_{\beta\dot\alpha}=A_{\alpha\dot\alpha}$.}
\begin{equation}\label{eq:FixingLambdaInADef}
a^\star_{\dot\alpha}(x,\lambda)=A^\star_{\alpha\dot\alpha}(x)\star\lambda^\alpha.
\end{equation}
Hence, the gauge fixing makes the dependence of the fields on the fiber coordinates explicit via equations (\ref{eq:GaugeFixingFieldsBFDef}) and (\ref{eq:FixingLambdaInADef}). This in principle allows us to integrate out the fiber $\mathbb{CP}^1$ on the action $S_1^\star$. However, the dependence on the $\mathbb{CP}^1$ is not star-commutative. In order to make sense of the integration over the fiber, we need to star commute every $\lambda$ and $\hat\lambda$ close to the measure on the $\mathbb{CP}^1$. For this purpose, we will explicitly compute the gauge fixed field strength two-form $\bar\partial a + a\wedge a$ in order to extract the dependence on the fiber:
\begin{align}\label{dA+AA}
\begin{split}
&(\bar\partial a + a\wedge a)\big|_{\tiny \mbox{gauge fixed}}\\
=\ & \bar\partial\left(A^\star_{\beta\dot\beta}\star\lambda^\beta\star\bar e^{\dot\beta}\right) + \big(A^\star_{\alpha\dot\alpha}\star\lambda^\alpha\star\bar e^{\dot\alpha}\big) \wedge_\star \big(A^\star_{\beta\dot\beta}\star\lambda^\beta\star\bar e^{\dot\beta}\big)\\
=\ & \Big(\partial^\star_{\alpha\dot\alpha}A^\star_{\beta\dot\beta} + A^\star_{\gamma\dot\gamma}\star R_{\alpha}{}^{\gamma}R_{\dot\alpha}{}^{\dot\gamma}A^\star_{\beta\dot\beta} \Big)\star (\lambda^\alpha\star\bar e^{\dot\alpha})\wedge_\star(\lambda^\beta\star\bar e^{\dot\beta})\\
=\ & G^\star_{\alpha\dot\alpha\beta\dot\beta} \star (\lambda^\alpha\star\bar e^{\dot\alpha})\wedge_\star(\lambda^\beta\star\bar e^{\dot\beta}),
\end{split}
\end{align}
where in the third line we used that, due to our gauge choice in equation \eqref{eq:gaugeCondStar}, $\bar\partial_0(A^\star_{\alpha\dot\alpha}\star\lambda^\alpha)=0$. Additionally, we introduce the ${\cal R}$-matrix factors in the appropriate representation when commuting $ \lambda^\alpha$ and $\bar e^{\dot\alpha}$ with $A^\star_{\beta\dot\beta}$. In the last step we recognized the component of the field strength two-form in terms of the star product basis from \eqref{eq:gaugeTwoComponent}. This is further simplified by using the identity
\begin{equation}
(\bar e^{\dot\alpha}\star\lambda^\alpha)\wedge_\star(\bar e^{\dot\beta}\star\lambda^\beta)=\tensor{R}{_\gamma^\alpha_{\dot\gamma}^{\dot\beta}}\bar e^{\dot\alpha}\wedge_\star\bar e^{\dot\gamma}\star\lambda^\gamma\star\lambda^\beta=\tensor{R}{_\gamma^\alpha_{\dot\gamma}^{\dot\beta}}\epsilon^{\dot\alpha\dot\gamma}\bar e^{\dot\delta}\wedge\bar e_{\dot\delta}(\lambda^\gamma\star\lambda^\beta).
\end{equation}
Consequently, the coefficient of (\ref{dA+AA}) is the component of the anti self-dual field strength:
\begin{equation}
\begin{aligned}
(\bar\partial a + a\wedge a)\Big|_{\tiny \mbox{gauge fixed}} &= G^\star_{\alpha\dot\alpha\beta\dot\beta}
\tensor{R}{_\gamma^\alpha_{\dot\gamma}^{\dot\beta}}\epsilon^{\dot\alpha\dot\gamma}\star(\lambda^\gamma\star\lambda^\beta)\star\left(\bar e^{\dot\delta}\wedge\bar e_{\dot\delta}\right)\\
&= G^\star_{\alpha\beta}\star(\lambda^\alpha\star\lambda^\beta)\star\left(\bar e^{\dot\delta}\wedge\bar e_{\dot\delta}\right)
\end{aligned}
\end{equation}
Inserting this into the action (\ref{BF*}), results in
\begin{align}
\begin{split}
S_1^\star[a,B] = & \frac{i}{2\pi}\int \frac{\Omega^3\wedge\bar\Omega^3}{\langle\lambda\hat\lambda\rangle^4}\wedge\tr\left[3B_{\rho\sigma}^\star\star \frac{\hat\lambda^\rho\star\hat\lambda^\sigma}{\langle\lambda\hat\lambda\rangle^2} \wedge_\star G^\star_{\alpha\beta}\star(\lambda^\alpha\star\lambda^\beta) \right]\\
= & \frac{i}{2\pi}\int \frac{\Omega^3\wedge\bar\Omega^3}{\langle\lambda\hat\lambda\rangle^4}\wedge\tr\left[ 3B_{\rho\sigma}^\star\star R_\gamma{}^\rho R_\delta{}^\sigma G^\star_{\alpha\beta}\star\frac{\hat\lambda^\gamma\star\hat\lambda^\delta}{\langle\lambda\hat\lambda\rangle^2}\star(\lambda^\alpha\star\lambda^\beta) \right],
\end{split}
\end{align}
where in the first line, $\bar e^0$ is star commutative and not affected by the $\star$-product similar to $\bar e^{\dot\delta}\wedge\bar e_{\dot\delta}$ which together forms $\bar\Omega^3/\langle\lambda\hat\lambda\rangle^4$, which is star commutative as well. Therefore it can with $\Omega^3$ to the full measure on $\mathbb{CP}^3$. As a last step, we commute the remaining powers of fiber coordinates to the same place. Now, we will integrate over the fiber. First, we decompose the twistor space measure using equation (\ref{eq:SplitVol}):
\begin{align}\label{3stepS}
S_1^\star[a,B] = & \frac{i}{2\pi}\int \mathrm{d}^4x\ \tr\left[ B_{\rho\sigma}^\star\star R_\gamma{}^\rho R_\delta{}^\sigma G^\star_{\alpha\beta}\left( \int_{\mathbb{CP}^1}\frac{\omega\wedge\bar\omega}{\langle\lambda\hat\lambda\rangle^2}\frac{\hat\lambda^\gamma\star\hat\lambda^\delta\star\lambda^\alpha\star\lambda^\beta}{\langle\lambda\hat\lambda\rangle^2}\right)\right],
\end{align}
where we used that $\omega$ and $\bar\omega$ are not affected by the $\star$-product. Finally, we integrate over the $\mathbb{CP}^1$ fibre. Using the cyclicity of the star product under integration and equation (\ref{l*l}), we have
\begin{equation}\label{eq:integrateCP1}
\begin{aligned}
   \int_{\mathbb{CP}^1}\frac{\omega\wedge\bar\omega} {\langle\lambda\hat\lambda\rangle^4}\hat\lambda^\gamma\star\hat\lambda^\delta\star\lambda^\alpha\star\lambda^\beta&=\int_{\mathbb{CP}^1}\frac{\omega\wedge\bar\omega} {\langle\lambda\hat\lambda\rangle^4}\left(\hat\lambda^\gamma\star\hat\lambda^\delta\right)\left(\lambda^\alpha\star\lambda^\beta\right)\\
   &=\int_{\mathbb{CP}^1}\frac{\omega\wedge\bar\omega} {\langle\lambda\hat\lambda\rangle^4}\left(\hat\lambda^\gamma\hat\lambda^\delta\right)\left(\lambda^\alpha\lambda^\beta\right)=-\frac{2\pi i}{3!}\left(\epsilon^{\gamma\beta}\epsilon^{\delta\alpha} + \epsilon^{\gamma\alpha}\epsilon^{\delta\beta}\right).
\end{aligned}
\end{equation}
Hence, the twistor space action reduces to
\begin{equation}
S_1^\star[a,B] = \frac{1}{2}\int \mathrm{d}^4x\ \tr\left[ B_{\rho\sigma}^\star\star R_\gamma{}^\rho R_\delta{}^\sigma G^\star_{\alpha\beta}\right]\left(\epsilon^{\gamma\beta}\epsilon^{\delta\alpha} + \epsilon^{\gamma\alpha}\epsilon^{\delta\beta}\right),
\end{equation}
which is the Anti-Self-Dual part of the twisted Yang-Mills theory, given by the first term of the action (\ref{noncomutativeaction}):
\begin{equation}\label{S1*=BG}
S_1^\star[a,b] = \int \mathrm{d}^4x\  \tr\left[B_{}\wedge_\star G_{_{ASD}}\right].
\end{equation}

We have shown that the noncommutative BF theory reproduces the ASD sector of the twisted Yang–Mills theory. It remains to prove that the additional non-local term in the twistor action (\ref{eq:FullTwistorAction})
\begin{equation}
S_2^\star[b] = -\frac{g^2_{\tiny \mbox{YM}}}{2}\int\ \tr\left[B_{}\wedge_\star B_{}\right].
\end{equation}
describes the additional part missing to the full description of the twisted YM theory in \eqref{noncomutativeaction}. To prove this, we will follow the same procedure as before, i.e., we use the gauge fixing form of the field (\ref{A0=0}) and then integrate over the fibre. First, we use the invariance of the one-form  $\omega_i$ under $\star$-product to write the non-local action as:
\begin{equation}
S_2^\star[B] = \frac{g^2_{\tiny \mbox{YM}}}{8\pi^2}\int \mathrm{d}^4x \int_{\mathbb{CP}^1\times \mathbb{CP}^1}\langle\lambda_1\lambda_2\rangle^2 \tr\Big[(\omega_1\wedge b_1)\star(\omega_2\wedge b_2)\Big].
\end{equation}
The integration over the $\mathbb{CP}^1$ fibre is only non-vanishing for the zeroth component of $B$. Therefore, in the action, we should make the following substitution:
\begin{equation}\label{subst}
\omega_i\wedge b_i \to \frac{\omega_i\wedge\bar\omega_i}{\langle\lambda_i\hat\lambda_i\rangle^2}3B^\star_{\alpha\beta}\star\frac{\hat\lambda^\alpha_i\star\hat\lambda^\beta_i}{\langle\lambda_i\hat\lambda_i\rangle^2}.
\end{equation}
Using this and that $\langle\lambda_1\lambda_2\rangle$ is not affected by the $\star$-product, we move this term to the middle of the action. After using the substitution (\ref{subst}) above, and using equation (\ref{l*l}), the action reads:
\begin{equation}
\begin{split}
&S_2^\star[b] =\frac{g^2_{\tiny \mbox{YM}}}{8\pi^2}\int \mathrm{d}^4x \int \tr\Big[(\omega_1\wedge b_1)\star\langle\lambda_1\lambda_2\rangle^2\star(\omega_2\wedge b_2)\Big]\\
& =\frac{g^2_{\tiny \mbox{YM}}}{8\pi^2}\int \mathrm{d}^4x\ \tr\int \text{vol}^{1,2}_{\mathbb{CP}^1} \left(3B^\star_{\alpha\beta}\star\frac{\hat\lambda^{\alpha}_1\star\hat\lambda^{\beta}_1}{\langle\lambda_1\hat\lambda_1\rangle^2}\right)\star(\lambda_{1\alpha'}\star\lambda_{1\beta'}) \star (\lambda_{2}^{\alpha'}\star\lambda_{2}^{\beta'})\star \left(3B^\star_{\gamma\delta}\star\frac{\hat\lambda^{\gamma}_2\star\hat\lambda^{\delta}_2}{\langle\lambda_2\hat\lambda_2\rangle^2}\right),
\end{split}
\end{equation}
where $\text{vol}^{i}_{\mathbb{CP}^1}=\frac{\omega_i\wedge\bar \omega_i}{\Braket{\lambda_i\hat\lambda_i}^2}$ denotes the volume form on $\mathbb{CP}^1$ and we used the star-commutativity of Poincar\'e invariant objects such as $\Braket{\lambda_i\lambda_j}$ and of the volume forms. Using the identity in \eqref{eq:integrateCP1} for integrating over both $\mathbb{CP}^1$, the latter becomes
\begin{align}
\begin{split}
S_2^\star[b]&=-\frac{g^2_{\tiny \mbox{YM}}}{2}\int \mathrm{d}^4x\ \tr\left[  B^\star_{\alpha\beta}\left(\delta^{\alpha}_{\alpha'}\delta^{\beta}_{\beta'}+\delta^{\alpha}_{\beta'}\delta^{\beta}_{\alpha'}\right)\star R_{\gamma'}{}^{\alpha'}R_{\delta'}{}^{\beta'}B^\star_{\gamma\delta}\left(\epsilon^{\alpha'\gamma}\epsilon^{\beta'\delta} + \epsilon^{\alpha'\delta}\epsilon^{\beta'\gamma}\right)\right]\\
&= -\frac{g^2_{\tiny \mbox{YM}}}{2}\int \mathrm{d}^4x \ \tr\left[ B_{\alpha\beta}^\star\star R_\alpha{}^\gamma R_\beta{}^\delta B^\star_{\gamma\delta}\right]\\
&= -\frac{g^2_{\tiny \mbox{YM}}}{2}\int \tr\big( B_{}\wedge_\star B_{}\big).
\end{split}
\end{align}
Hence, by imposing the same incidence relation on the twisted twistor space as in the undeformed case, the gauge fixed noncommutative twistor action reproduces the full noncommutative Yang-Mills action in four-dimensional spacetime. This proof does not immediately generalize to twists chosen from the full conformal algebra, as it relies on certain objects to remain star commutative.

\subsection{Extension to noncommutative deformations of $\mathcal{N}=4$ SYM}

In the following, we will discuss how the latter construction of gauge theories on twisted twistor space can be extended to supertwistor space. Since the twists we are considering deform only the Poincar\'e algebra, which acts trivially on the fermionic coordinates $\psi^i$, the fermionic supercoordinates will be star-commutative. For constructions involving deformations of the $SU(4)$ R-symmetry, cf. \cite{Adamo:fish}.

As in the purely bosonic twistor space studied in the first part of this section, we keep the the incidence relation identical to the undeformed case. The additional super-incidence relation likewise remains intact and can, in turn, be written in terms of the star product:
\begin{equation}
\psi^i = \theta^{i\alpha}\lambda_\alpha = \theta^{i\alpha}\star\lambda_\alpha.
\end{equation}
Note that Poincar\'e transformations act similarly on $\theta^i_\alpha$ as on $\lambda_\alpha$ and hence, the star product in the latter expression trivializes. The projection of the fermionic variable is then also not affected by the star product:
\begin{equation}
\theta^{i\alpha} = \frac{\hat\psi^i\star\lambda^\alpha - \psi^i\star\hat\lambda^\alpha}{\braket{\lambda\stackrel{\star}{,}\hat\lambda}} = \frac{\hat\psi^i\lambda^\alpha - \psi^i\hat\lambda^\alpha}{\langle\lambda\hat\lambda\rangle}.
\end{equation}

We can now consider an extension of the gauge theory studied in the last section, and define a noncommutative theory on the super-twistor space. This theory will then be shown to be perturbatively equivalent to the noncommutative version of the ${\cal N}=4$ SYM theory presented in \eqref{eq:ActionDefSYM}. Like in the pure Yang-Mills case in the previous subsection, the noncommutative deformation is obtained as a star-product version of the undeformed action described in section \ref{subsec:twistorSYM}. We therefore introduce a superfield ${\cal A}$, that now lives on the noncommutative version of the supertwistor space. We can also expand ${\cal A}$ order by order on the fermionic fields $\psi^i$, similar to equation (\ref{calA}). As a convenient notation, that will proven to be usefull in this section, we write:
\begin{align}\label{eq:AComps}
{\cal A} = a + \tilde{\cal X} + \hat\Phi + {\cal X} + {\cal B}
\end{align}
where:
\begin{align}\label{eq:components}
\tilde{\cal X}  = \tilde\chi_i\psi^i,\ \ \ \ \hat\Phi = \frac{1}{2}\phi_{ij}\psi^i\psi^j,\ \ \ \ {\cal X} = \frac{1}{3!}\chi^i\epsilon_{ijkl}\psi^j\psi^k\psi^l,\ \ \ \ {\cal B} = \frac{1}{4!}b\psi^4.
\end{align}

Beyond depending on the superspace coordinates $\psi$, the gauge field $\mathcal{A}$ is a (0,1)-form and hence, we choose to express it in terms of the star basis of differential forms as
\begin{align}
\begin{split}
&\hspace{3cm}a = a^\star_0\star \bar e^0 + a^\star_{\dot\alpha}\star\bar e^{\dot\alpha},\ \ \ \ b = b^\star_0\star \bar e^0 + b^\star_{\dot\alpha}\star\bar e^{\dot\alpha}\\
&\tilde\chi_i = \tilde\chi_{i,0}^\star\star\bar e^0 + \tilde\chi_{i\dot\alpha}^\star\star\bar e^{\dot\alpha},\ \ \ \chi^i = \chi^{i\star}_{0}\star\bar e^0 + \chi^{i\star}_{\dot\alpha}\star\bar e^{\dot\alpha},\ \ \ \ \phi_{ij} = \phi^\star_{ij,0}\star\bar e^0 + \phi^\star_{ij,\dot\alpha}\star\bar e^{\dot\alpha},
\end{split}
\end{align}
similarly as in \eqref{eq:YMFieldUsingStarProduct} for the pure Yang-Mills case.

Again, in the noncommutative setting, the ordinary gauge transformations are replaced by their noncommutative star-gauge transformations and hence the supergauge field $\mathcal{A}$ transforms as
\begin{equation}\label{eq:deltaAstar}
    \delta\mathcal{A}=\bar\partial\epsilon+\left[\mathcal{A}\stackrel{*}{,}\epsilon\right].
\end{equation}

\paragraph{Gauge invariant action}

As in the undeformed case, the twist noncommutative supertwistor theory will be defined as a sum of two different terms. The first term will be given by a noncommutative version of the holomorphic Chern-Simons action on $\mathbb{CP}^{3|4}$, and will be responsible for describing the ASD part of the noncommutative ${\cal N}=4$ SYM theory. The other term is the deformation of the non-local term discussed in section \ref{subsec:twistorSYM} which will complete the deformed $\mathcal{N}=4$ SYM:
\begin{equation}\label{eq:DefSuperTwistorAction}
\begin{aligned}
S^\star[{\cal A}] &= S^\star_1[{\cal A}]+S^\star_2[{\cal A}]\\
S^\star_1[{\cal A}]&=\int \Omega^{3|4}\wedge_\star \tr\left({\cal A}\wedge_\star \bar\partial{\cal A} + \frac{2}{3}{\cal A}\wedge_\star {\cal A}\wedge_\star {\cal A} \right)\\
S_2^\star[{\cal A}] &= -g^2_{\tiny\mbox{YM}}\sum_{n=2}^\infty\frac{1}{n}\left(\frac{1}{2\pi i}\right)^n\int_{(\mathbb{CP}^1)^n} \frac{\omega_1\cdots\omega_n}{\langle\lambda_1\lambda_2\rangle\cdots\langle\lambda_n\lambda_1\rangle}\tr\left({\cal A}_1\star\cdots\star{\cal A}_n\right).
\end{aligned}
\end{equation}

The twisted hCS action transforms under gauge transformations as:
\begin{equation}
\delta_\epsilon S^\star_1[{\cal A}] = \tr\int\Big[\epsilon\stackrel{\star}{,}\Omega^{3|4}\wedge_\star\big({\cal A}\wedge_\star \bar\partial{\cal A} + \frac{2}{3}{\cal A}\wedge_\star {\cal A}\wedge_\star {\cal A}\big) \Big].
\end{equation}
Note that the holomorphic top form $\Omega^{3|4}$ is invariant under the full conformal algebra and hence is star commutative. Now, the cyclicity of the star product under integration implies that t$S_1^\star$ is gauge invariant. As in the undeformed case, the gauge freedom \eqref{eq:deltaAstar} will be used to impose a gauge fixing condition. We can express the super-one-form in components as ${\cal A} = {\cal A}^\star_0\star\bar e^0 + {\cal A}_{\dot\alpha}^\star\star\bar e^{\dot\alpha}$ and impose the noncommutative equivalent of the \textit{Woodhouse} gauge condition:
\begin{equation}\label{eq:WoodhouseGaugeDef}
\bar\partial^*\big|_X{\cal A}_0^\star = 0,
\end{equation}
where $\bar\partial^*|_X$ is the adjoint of the $\bar\partial$-operator restricted to any twistor line $X=\mathbb{CP}^1$. The condition \eqref{eq:WoodhouseGaugeDef} does not completely fixes the gauge freedom \eqref{eq:deltaAstar}, and the residual gauge transformations are harmonic functions on the fibres of twistor space \cite{Mason:twist,Adamo:fish}:
\begin{equation}\label{eq:ResidualGauge}
\bar\partial^*\big|_X\bar\partial\big|_X\epsilon(Z) = 0.
\end{equation}
The only degree-zero homogeneous harmonic functions on $X$ are constants. Then, the residual gauge freedom left unfixed by \eqref{eq:WoodhouseGaugeDef} is precisely the spacetime gauge transformations $\epsilon(Z)=\epsilon(x)$, like in the undeformed case.

The gauge condition \eqref{eq:WoodhouseGaugeDef} must hold for each component of the $\psi$-expansion separately,
\begin{equation}
\bar\partial^*\big|_{X}a^\star_0 = \bar\partial^*\big|_{X}\tilde\chi^\star_0 = \bar\partial^*\big|_{X}\phi^\star_0 = \bar\partial^*\big|_{X}\chi^\star_0 = \bar\partial^*\big|_{X}b^\star_0=0.
\end{equation}
By the same argument performed in the undeformed case, we can see that this gauge fixing leads to the following form of the component fields:
\begin{align}\label{eq:GaugeFixedSYMDef}
\begin{split}
a = a^\star_{\dot\alpha}\star\bar e^{\dot\alpha},&\ \ \ \ \tilde\chi_i = \tilde\chi^\star_{i\dot\alpha}\star\bar e^{\dot\alpha},\\
\phi_{ij} = \Phi_{ij}^\star\star\bar e^0+\varphi^\star_{ij,\dot\alpha}\star\bar e^{\dot\alpha},&\ \ \ \ \chi^i = 2\frac{\hat\lambda^\alpha}{\langle\lambda\hat\lambda\rangle}\star\Psi^{i\star}_{\alpha}\star\bar e^0 + \psi^{i\star}_{\dot\alpha}\star\bar e^{\dot\alpha},\\
b = 3B^\star_{\alpha\beta}\star\frac{\hat\lambda^\alpha\star\hat\lambda^\beta}{\langle\lambda\hat\lambda\rangle^2}\star\bar e^0 & +b_{\dot\alpha}^\star\star\bar e^{\dot\alpha},
\end{split}
\end{align}
where the component fields $\Phi^\star_{ij}, \Psi^{i\star}_\alpha, B^\star_{\alpha\beta}$ depend only on $x$. For later use and clarity, we introduce a more compact form that combines the component fields defined in the above expressions with the appropriate basis forms, such that 
\begin{align}
\begin{split}
\Phi_{0,ij} = \Phi_{ij}^\star\star\bar e^0&,\ \ \ \ \varphi_{ij}= \varphi^\star_{ij,\dot\alpha}\star\bar e^{\dot\alpha},\\
\Psi^i_0 = \frac{\hat\lambda^\alpha}{\langle\lambda\hat\lambda\rangle}\star\Psi^{i\star}_{\alpha}\star\bar e^0&,\ \ \ \ \xi^i = \psi^{i\star}_{\dot\alpha}\star\bar e^{\dot\alpha},\\
b_0=3B^\star_{\alpha\beta}\star\frac{\hat\lambda^\alpha\star\hat\lambda^\beta}{\langle\lambda\hat\lambda\rangle^2}\star\bar e^0&,\ \ \ \ \hat b=B_{\dot\alpha}^\star\star\bar e^{\dot\alpha}.
\end{split}
\end{align}

The gauge invariance of the non-local action $S_2^\star$ is explained in Appendix \ref{app:GaugeInvarianceNonLocal}. Using the gauge-fixed form of the superfield ${\cal A}$ in \eqref{eq:GaugeFixedSYMDef} and expanding the action $S_2^\star$ in powers of the fermionic coordinates $\psi^I$, only a finite number of contributions appear, which are given as
\begin{align}\label{eq:NonLocalDefSYMGaugeFix}
\begin{split}
S^\star_2[{\cal A}]\Big|_{\tiny\mbox{gauge fixed}}&=-\frac{g^2_{\tiny\mbox{YM}}}{2(2\pi i)^2} \int d^{4|8}x \int_{(\mathbb{CP}^1)^2} \frac{\omega_1\omega_2}{\langle\lambda_1\lambda_2\rangle\langle\lambda_2\lambda_1\rangle}\tr\left({\cal B}_1\star {\cal B}_2\right)\\
&-\frac{g^2_{\tiny\mbox{YM}}}{3(2\pi i)^3} \int d^{4|8}x\int_{(\mathbb{CP}^1)^3} \frac{\omega_1\omega_2\omega_3}{\langle\lambda_1\lambda_2\rangle\langle\lambda_2\lambda_3\rangle\langle\lambda_3\lambda_1\rangle}\tr\left( {\cal X}_1\star\hat\Phi_2\star{\cal X}_3 \right) \\
&-\frac{g^2_{\tiny\mbox{YM}}}{4(2\pi i)^4}\int d^{4|8}x \int_{(\mathbb{CP}^1)^4} \frac{\omega_1\omega_2\omega_3\omega_4}{\langle\lambda_1\lambda_2\rangle \langle\lambda_2\lambda_3\rangle \langle\lambda_3\lambda_4\rangle \langle\lambda_4\lambda_1\rangle}\tr\left( \hat\Phi_1\star\hat\Phi_2\star\hat\Phi_3\star\hat\Phi_4 \right).
\end{split}
\end{align}

\paragraph{Integrating the fibre}

So far, we have constructed a noncommutative version of the Chern-Simons theory on the super-twistor space together with a non-local term, as given by the action (\ref{eq:DefSuperTwistorAction}). The gauge fixing procedure explained above will be the intermediate step towards the integration over the line bundle $\mathbb{CP}^1$, which as a result will recover the four-dimensional spacetime theory. We will start with the local part of the action. Applying the gauge fixing (\ref{eq:GaugeFixedSYMDef}) to the action and performing the fermionic integral results in
\begin{align}
    S_1^\star[{\cal A}] = \int\  & \Omega^3\wedge_\star \tr\Big[ b_0\wedge_\star\left(\bar\partial a + [a\wedge_\star a]\right) + \hat b\wedge_\star\bar\partial a \Big] \tag{I} \label{eq:LN=4I}\\
+ & \Omega^3\wedge_\star \tr\Big[ \varphi^{ij}\wedge_\star\left(\bar\partial\Phi_{0,ij} + [a\wedge_\star\Phi_{0,ij}]\right) + \varphi^{ij}\wedge_\star\bar\partial\varphi_{ij} \Big] \quad \tag{II}\label{eq:LN=4II}\\
+ & \Omega^3\wedge_\star\tr\Big[ \tilde\chi_i\wedge_\star\left(\bar\partial\Psi^i_0 + [a\wedge_\star\Psi^i_0]\right) + \xi^i\wedge_\star\bar\partial\tilde\chi_i \Big] \tag{III}\label{eq:LN=4III}\\
+ &  \Omega^3\wedge_\star\tr\left[ \tilde\chi_i \wedge_\star \left(\Phi^{ij}_{0}\wedge_\star \tilde\chi_j \right)\right]. \tag{IV}\label{eq:LN=4IV}
\end{align}
The fields $\hat b$ and $\xi^i$ appear only linearly; therefore, they play the role of Lagrange multipliers. Integrating them out results in
\begin{equation}
\begin{aligned}
\bar\partial_0 a^\star_{\dot\alpha} &=0,\\
\bar\partial_0 \tilde \chi_{i\dot\alpha}^\star &=0,
\end{aligned}
\end{equation}
respectively. Since $a^\star_{\dot\alpha}$ and $\chi_{i\dot\alpha}^\star$ have homogeneous weights $+1$ and $0$, respectively, this results in
\begin{equation}\label{gfixn=4}
a^\star_{\dot\alpha} = A^\star_{\alpha\dot\alpha}(x)\star\lambda^\alpha,\ \ \ \ \tilde\chi_{i\dot\alpha}^\star = \bar\Psi_{i\dot\alpha}^\star(x).
\end{equation}
The expression \eqref{eq:LN=4I} is identical to the action studied in the last section and therefore will result in the ASD part of the pure Yang-Mills action:
\begin{equation}
\eqref{eq:LN=4I} = \int_{\mathbb{R}^4}\tr\Big(B_{}\wedge_\star G_{_{ASD}}\Big).
\end{equation}
In the second term \eqref{eq:LN=4II}, the field $\varphi_{ij}$ appears quadratically and therefore it is possible to use its equations of motion:
\begin{align}
\begin{split}
\eqref{eq:LN=4II}=\ &\varphi^{ij}\wedge_\star\left(\bar\partial\Phi_{0,ij} + [a\wedge_\star\Phi_{0,ij}]\right) + \varphi^{ij}\wedge_\star\bar\partial\varphi_{ij}\\
=\ &\varphi^{ij\star}_{\dot\alpha}\star\bar e^{\dot\alpha}\wedge_\star \left( \bar\partial^\star_{\dot\beta}\Phi_{ij}\star\bar e^{\dot\beta}\wedge_\star\bar e^0 + [a^\star_{\dot\beta}\star\bar e^{\dot\beta}\wedge_\star \Phi_{ij}\star\bar e^0] \right) + \varphi^{ij\star}_{\dot\alpha}\star\bar e^{\dot\alpha}\wedge_\star\bar\partial_0\varphi_{ij,\dot\beta}^\star\star\bar e^0\wedge_\star\bar e^{\dot\beta}\\
=\ & \varphi^{ij\star}_{\dot\delta}\star R_{\dot\alpha}{}^{\dot\delta}\Big[ (\bar\partial^\star_{\dot\beta}\Phi_{ij} + a^\star_{\dot\gamma}\star R_{\dot\beta}{}^{\dot\gamma}\Phi_{ij}) + \bar\partial_0\varphi^{\star}_{ij,\dot\beta} \Big]\epsilon^{\dot\alpha\dot\beta}\frac{\bar\Omega}{\langle\lambda\hat\lambda\rangle^4}\\
=\ & \varphi^{ij\star}_{\dot\delta}\star R_{\dot\alpha}{}^{\dot\delta}\Big[ (\partial_{\beta\dot\beta}^\star\Phi_{ij}  + A^\star_{\gamma\dot\gamma}\star R_{\beta}{}^{\gamma}R_{\dot\beta}{}^{\dot\gamma}\Phi_{ij})\star\lambda^\beta + \bar\partial_0\varphi^{\star}_{ij,\dot\beta} \Big]\epsilon^{\dot\alpha\dot\beta}\frac{\bar\Omega}{\langle\lambda\hat\lambda\rangle^4},
\end{split}
\end{align}
Where we used (\ref{gfixn=4}) in the last line. Integrating out the field $\varphi^\star_{ij,\dot\alpha}$, we get the equation of motion:
\begin{equation}
\bar\partial_0\varphi^\star_{ij,\dot\alpha} = (\partial_{\alpha\dot\alpha}^\star\Phi_{ij}  + A^\star_{\gamma\dot\gamma}\star R_{\alpha}{}^{\gamma}R_{\dot\alpha}{}^{\dot\gamma}\Phi_{ij})\star\lambda^\alpha = D^\star_{\alpha\dot\alpha}\Phi_{ij}\star\lambda^\alpha,
\end{equation}
and hence,
\begin{equation}\label{solphi}
\varphi^\star_{ij,\dot\alpha} = D^\star_{\alpha\dot\alpha}\Phi_{ij}\star\frac{\hat\lambda^\alpha}{\langle\lambda\hat\lambda\rangle}.
\end{equation}
We can now plug (\ref{gfixn=4}) and (\ref{solphi}) into the action, and hence \eqref{eq:LN=4II} becomes
\begin{align}
\begin{split}
\eqref{eq:LN=4II}=\ &\int_{\mathbb{CP}^3}\tr\ \frac{1}{\langle\lambda\hat\lambda\rangle} (D_{\alpha\dot\alpha}^\star\Phi_{ij}\star\hat\lambda^\alpha)\star R_{\dot\gamma}{}^{\dot\alpha}(D^\star_{\beta\dot\beta}\Phi^{ij})\epsilon^{\dot\gamma\dot\beta}\star\lambda^\beta \ \frac{\bar\Omega\wedge\Omega}{\langle\lambda\hat\lambda\rangle^4}\\
=\ &\int_{\mathbb{CP}^3}\tr\ (D_{\alpha\dot\alpha}^\star\Phi_{ij})\star R_{\gamma}{}^{\alpha}R_{\dot\gamma}{}^{\dot\alpha}(D^\star_{\beta\dot\beta}\Phi^{ij})\epsilon^{\dot\gamma\dot\beta}\star \frac{\hat\lambda^{\gamma}\star\lambda^{\beta}}{\langle\lambda\hat\lambda\rangle} \ \frac{\bar\Omega\wedge\Omega}{\langle\lambda\hat\lambda\rangle^4}\\
=\ &\int_{\mathbb{R}^4}\tr\ (D_{\alpha\dot\alpha}^\star\Phi_{ij})\star R_{\gamma}{}^{\alpha}R_{\dot\gamma}{}^{\dot\alpha}(D^\star_{\beta\dot\beta}\Phi^{ij})\epsilon^{\dot\gamma\dot\beta}\star \int_{\mathbb{CP}^1} \frac{\hat\lambda^{\gamma}\star\lambda^{\beta}}{\langle\lambda\hat\lambda\rangle} \ \frac{\bar\omega\wedge\omega}{\langle\lambda\hat\lambda\rangle^2}\ \mathrm{d}^4x \\
=\ &\int_{\mathbb{R}^4}\tr\ (D_{\alpha\dot\alpha}^\star\Phi_{ij})\star R_{\gamma}{}^{\alpha}R_{\dot\gamma}{}^{\dot\alpha}(D^\star_{\beta\dot\beta}\Phi^{ij})\epsilon^{\dot\gamma\dot\beta}\epsilon^{\gamma\beta}\mathrm{d}^4x
\end{split}
\end{align}
In the latter, several R matrices appear in appropriate representations resulting from commuting several $\lambda^\alpha$ and basis forms $\bar e^{\dot\alpha}$ over several fields. Furthermore, we use the formula \eqref{eq:IntegrateOut} to integrate over the fibre, resulting in the kinetic term for the scalar field for the noncommutative gauge theory. Written in terms of differential forms, this leads to
\begin{equation}
\eqref{eq:LN=4II} = \int_{\mathbb{R}^4} \tr\left(D\Phi_{ij}\wedge_\star *D\Phi^{ij}\right).
\end{equation}
Similarly, \eqref{eq:LN=4IV} leads to the right-handed part of the Yukawa interactions as
\begin{align}
\begin{split}
\eqref{eq:LN=4IV}=\ &\int_{\mathbb{CP}^3}\Omega^3\wedge\tr\Big( \tilde\chi_i \wedge_\star [\Phi^{ij}_0\wedge_\star \tilde\chi_j] \Big)\\
=\ & \int_{\mathbb{CP}^3}\Omega^3\wedge\tr \Big( \bar\Psi^\star_{i\dot\alpha}\star\bar e^{\dot\alpha}\wedge_\star[ \Phi^{ij}\star\bar e^0 \wedge_\star \bar\Psi^\star_{j\dot\beta}\star\bar e^{\dot\beta}] \Big)\\
=\ & \int_{\mathbb{CP}^3}\tr \Big( \bar\Psi^\star_{i\dot\alpha}\star [R_{\dot\delta}{}^{\dot\gamma} \Phi^{ij}\star R_{\dot\gamma}{}^{\dot\beta}\bar\Psi^\star_{j\dot\beta}] \Big)\epsilon^{\dot\alpha\dot\delta} \ \frac{\Omega^3\wedge\bar\Omega^3}{\langle\lambda\hat\lambda\rangle^4}\\
=\ & \int_{\mathbb{R}^4}\tr \Big( \bar\Psi^\star_{i\dot\alpha}\star [R_{\dot\delta}{}^{\dot\gamma} \Phi^{ij}\star R_{\dot\gamma}{}^{\dot\beta}\bar\Psi^\star_{j\dot\beta}] \Big)\epsilon^{\dot\alpha\dot\delta}\ \mathrm{d}^4x
\end{split}
\end{align}
To write the interaction term above index-free, we introduce the basis elements $\bar s^{\dot\alpha}$ as introduced in equation (\ref{eq:HalfFormsPsiDeformed}). Using the relations (\ref{eq:RForSpinorBasis}) to commute the basis element $\bar s^{\dot\alpha}$ in order to combine to the index free fields, results in
\begin{equation}
\eqref{eq:LN=4IV} = \int d^2\bar s \int_{\mathbb{R}^4}\mathrm{d}^4x\ \tr\Big( \bar\Psi_i \star [\Phi^{ij}\star\bar\Psi_j] \Big).
\end{equation}

For the term \eqref{eq:LN=4III}, we need to be more careful, since the interaction between the deformed covariant derivative $D^\star_{\alpha\dot\alpha}$ and the spinorial indices is non-trivial. To deal with it, we consider the object $(D_{\alpha\dot\alpha}\Psi_\beta)^\star$, defined by the action of $D^\star_{\alpha\dot\alpha}$ on the half-form spinor field $\Psi=s^\beta\star\Psi^\star_\beta$:
\begin{equation}\label{eq:CovDerOnSpinors}
D^\star_{\alpha\dot\alpha}(\Psi) = D^\star_{\alpha\dot\alpha}(s^\beta\star\Psi_\beta^\star) \equiv s^{\beta}\star(D_{\alpha\dot\alpha}\Psi_\beta)^\star.
\end{equation}
Since $\lambda^\alpha$ behaves similarly under Poincar\'e generators on twistor space as the basis spinor $s^\alpha$ does, the latter expression also holds for replacing $s$ by $\lambda$ and hence
\begin{align}
\begin{split}
\bar\partial\Psi^i_{0} + [a\wedge_\star\Psi^i_0] =\ & \frac{1}{\langle\lambda\hat\lambda\rangle}\left[\bar\partial\left(\hat\lambda^\beta\star\Psi^{i\star}_\beta\star\bar e^0\right)+ \Big( A_{\alpha\dot\alpha}^\star\star\lambda^\alpha\star\bar e^{\dot\alpha}\Big)\wedge_\star\left(\hat\lambda^\beta\star\Psi^{i\star}_\beta\star\bar e^0\right) \right]\\
=\ & \frac{1}{\langle\lambda\hat\lambda\rangle}\left[\bar\partial^\star_{\alpha\dot\alpha}\left(\hat\lambda^\beta\star\Psi^{i\star}_\beta \right) + A^\star_{\gamma\dot\gamma}\star R_{\alpha}{}^{\gamma}R_{\dot\alpha}{}^{\dot\gamma}\left(\hat\lambda^\beta\star\Psi^{i\star}_\beta\right)\right]\star\lambda^\alpha\star\bar e^{\dot\alpha}\wedge_\star\bar e^0\\
=\ & \frac{1}{\langle\lambda\hat\lambda\rangle} D^\star_{\alpha\dot\alpha}\left(\hat\lambda^\beta\star\Psi^{i\star}_\beta \right)\star\lambda^\alpha\star\bar e^{\dot\alpha}\wedge_\star\bar e^0\\
=\ &  \frac{1}{\langle\lambda\hat\lambda\rangle} \hat\lambda^\beta\star \left(D_{\alpha\dot\alpha}\Psi^i_\beta\right)^\star\star\lambda^\alpha\star\bar e^{\dot\alpha}\wedge_\star\bar e^0\\
=\ & R_{\gamma}{}^{\beta}\left(D_{\alpha\dot\alpha}\Psi^i_\beta\right)^\star\star \frac{\hat\lambda^\gamma\star\lambda^\alpha}{\langle\lambda\hat\lambda\rangle}\star\bar e^{\dot\alpha}\wedge_\star\bar e^0,
\end{split}
\end{align}
where we used the gauge fixed form of the fields in the first line and in the second line we commute $\bar e^{\dot\alpha}$ and $\lambda^\alpha$ with $\hat\lambda^\beta\star\Psi_\beta^{i\star}$ by introducing two $R$-matrices. In the last two lines, we used the definition (\ref{eq:CovDerOnSpinors}) and then commute $\hat\lambda^\beta$ with $(D_{\alpha\dot\alpha}\Psi^i_\beta)^\star$. Thus, (III) becomes
\begin{align}\label{eq:IIIwithIndex}
\begin{split}
\eqref{eq:LN=4III} =\ &\int\Omega^3\wedge\tr\Big[ \tilde\chi_i\wedge_\star\left(\bar\partial\Psi^i_0 + [a\wedge_\star\Psi^i_0]\right) \Big]\\
= & \int\Omega^3\wedge\tr \left[ \bar\Psi^{\star}_{i\dot\beta}\star\bar e^{\dot\beta}\wedge_\star R_{\gamma}{}^{\beta}\left(D_{\alpha\dot\alpha}\Psi^i_\beta\right)^\star\star \frac{\hat\lambda^\gamma\star\lambda^\alpha}{\langle\lambda\hat\lambda\rangle}\star\bar e^{\dot\alpha}\wedge_\star\bar e^0 \right]\\
= & \int\Omega^3\wedge\tr \left[ \bar\Psi^\star_{i\dot\beta}\star R_{\dot\gamma}{}^{\dot\beta} R_{\gamma}{}^{\beta}\left(D_{\alpha\dot\alpha}\Psi^i_\beta\right)^\star\star\bar e^{\dot\gamma}\star \frac{\hat\lambda^\gamma\star\lambda^\alpha}{\langle\lambda\hat\lambda\rangle}\star\bar e^{\dot\alpha}\wedge_\star\bar e^0 \right],
\end{split}
\end{align}
where in the last line we introduce an $R$-matrix by commuting $\bar e^{\dot\beta}$ over $\left(D_{\alpha\dot\alpha}\Psi^i_\beta\right)^\star$. After commuting $\bar e^{\dot\gamma}$ with $\hat\lambda^\gamma$ and $\lambda^\alpha$, we can combine the objects properly and integrate over the fibre $\mathbb{CP}^1$. We will therefore get an integration over $\mathbb{R}^4$. We expand the result by inserting the grassmann valued basis spinors  introduced in section \ref{sec:MatterFields} together with the appropriate integration and use star-commutation relations such that
\begin{equation}\label{eq:IIIwithIndexS}
\eqref{eq:LN=4III}=\int d^2s d^2\bar s\int \tr\left[ \bar\Psi^\star_{i\dot\beta}\star R_{\dot\gamma}{}^{\dot\beta}R_{\gamma}{}^{\beta}(D_{\alpha\dot\alpha}\Psi^i_\beta )^\star\star\bar s^{\dot\gamma}\star s^{\gamma}\star s^{\alpha}\star \bar s^{\dot\alpha} \right] \mathrm{d}^4x.
\end{equation}
By reorganizing the latter using the commutation relations in \eqref{eq:RForSpinorBasis}, the latter is equivalent to
\begin{equation}\label{eq:IIIwithIndexFree}
\eqref{eq:LN=4III}=\int d^2s d^2\bar s\ \tr\Big(\bar\Psi_i \star \sigma \wedge_\star *D\Psi^i \Big),
\end{equation}
which is exactly the deformed kinetic term for the fermionic fields appearing in the deformed $\mathcal{N}=4$ SYM.

Overall, the noncommutative hCS action on the super-twistor space is given by:
\begin{align}\label{SYM*ASD}
\begin{split}
S^\star_{1}[{\cal A}] = &\int  \tr\left[B_{}\wedge_\star G_{_{ASD}}\right] +  \int \tr\left(D\Phi^{ij}\wedge_\star *D\Phi_{ij}\right) + \int d^2s d^2\bar s\ \tr\Big(\bar\Psi_i \star \sigma \wedge_\star *D\Psi^i \Big)\\
& + \int d^2\bar s \int\mathrm{d}^4x\ \tr\Big( \bar\Psi_i \star [\Phi^{ij}\star\bar\Psi_i] \Big),
\end{split}
\end{align}
which is the twist-noncommutative deformed equivalent of the action \eqref{eq:ASD-SYMAction}.

In order to obtain the complete deformed $\mathcal{N}=4$ SYM action, we turn our attention to the non-local terms. The quadratic part of the non-local action, i.e. the first line in (\ref{eq:NonLocalDefSYMGaugeFix}) is the term we studied for the pure YM case, in the last section. This can be easily seen in analogy to the argument given for the undeformed case and will not be repeated here. The second line will give the missing left-handed Yukawa interactions. To see this, we plug the gauge-fixed form of the fields (\ref{eq:GaugeFixedSYMDef}) in the second term of (\ref{eq:NonLocalDefSYMGaugeFix}):
\begin{equation}
-\frac{g^2_{\tiny\mbox{YM}}}{3(2\pi i)^3}\int d^{4|8}x \int \frac{\omega_1\omega_2\omega_3}{\langle\lambda_1\lambda_2\rangle\langle\lambda_2\lambda_3\rangle\langle\lambda_3\lambda_1\rangle}\tr\left(  (\psi^i_1\psi^j_1\psi^k_1)\Psi^l_0\epsilon_{ijlk}\star(\psi^m_2\psi^n_2)\Phi_{0,mn}\star(\psi^p_3\psi^q_3\psi^r_1)\Psi^s_0\epsilon_{pqrs}  \right).
\end{equation}
We then integrate over the supercoordinates and substitute the gauge-fixed expressions for the fields leading to
\begin{align}
\begin{split}
&-\frac{g^2_{\tiny\mbox{YM}}}{3(2\pi i)^3}\int \mathrm{d}^4x \int \frac{\omega_1\wedge\bar\omega_1}{\langle\lambda_1\hat\lambda_1\rangle^2} \int \frac{\omega_2\wedge\bar\omega_2}{\langle\lambda_2\hat\lambda_2\rangle^2} \int \frac{\omega_3\wedge\bar\omega_3}{\langle\lambda_3\hat\lambda_3\rangle^2}\ \tr\left(\frac{\hat\lambda_1^\alpha}{\langle\lambda_1\hat\lambda_1\rangle}\star\Psi^{i\star}_\alpha \star \Phi_{ij}\star \frac{\hat\lambda_3^\beta}{\langle\lambda_3\hat\lambda_3\rangle}\star \Psi_\beta^{j\star} \right)\langle\lambda_1\lambda_3\rangle\\
&= -g^2_{\tiny\mbox{YM}}\int \mathrm{d}^4x\ \tr\Big(R^{\gamma}{}_{\beta}\Psi^{i\star}_{\alpha}\star R^\delta{}_{\gamma}\Phi_{ij}\star\Psi^{j\star}_\delta\Big)\epsilon^{\alpha\beta}\\
&= -g^2_{\tiny\mbox{YM}}\int d^2s\ \int\mathrm{d}^4x\ \tr\Big(\Psi^i\star[\Phi_{ij}\star\Psi^j]\Big),
\end{split}
\end{align}
where $\Psi = s^\alpha\star\Psi^\star_\alpha$. The last line of the non-local action (\ref{eq:NonLocalDefSYMGaugeFix}) results in the quartic scalar interactions. It is explicitly given by:
\begin{equation}
\begin{aligned}
-\frac{g^2_{\tiny\mbox{YM}}}{4(2\pi i)^4}\int d^{4|8}x &\int \frac{\omega_1\omega_2\omega_3\omega_4}{\langle\lambda_1\lambda_2\rangle\langle\lambda_2\lambda_3\rangle\langle\lambda_3\lambda_4\rangle\langle\lambda_4\lambda_1\rangle}\times\\
&\times\tr\left(  (\psi^i_1\psi^j_1)\Phi_{0,ij}\star (\psi^k_2\psi^l_2)\Phi_{0,kl}\star (\psi^m_3\psi^n_3)\Phi_{0,mn}\star (\psi^p_4\psi^q_4)\Phi_{0,pq} \right).
\end{aligned}
\end{equation}
As in the undeformed case, we use the result of section 2.5 of \cite{Koster} to integrate over the $\mathbb{CP}^1$ fibres. The action above then reduces to the quartic interaction term:
\begin{equation}
\frac{g^2_{\tiny\mbox{YM}}}{4}\int \mathrm{d}^4x\ \tr\Big([\Phi^{ij}\star\Phi^{kl}]\star[\Phi_{ij}\star\Phi_{kl}]\Big).
\end{equation}

Thus, the non-local action (\ref{eq:NonLocalDefSYMGaugeFix}) is given by:
\begin{align}
\begin{split}
S_2^\star[{\cal A}] =& -\frac{g^2_{\tiny\mbox{YM}}}{2}\int~\tr[B\wedge B]-g^2_{\tiny\mbox{YM}}\int d^2s\ \int\mathrm{d}^4x\ \tr\Big(\Psi^i\star[\Phi_{ij}\stackrel{\star}{,}\Psi^j]\Big)\\
&\hspace{2cm}+ \frac{g^2_{\tiny\mbox{YM}}}{4}\int\mathrm{d}^4x\ \tr\Big( [\Phi^{ij}\stackrel{\star}{,}\Phi^{kl}]\star[\Phi_{ij}\stackrel{\star}{,}\Phi_{kl}]\Big)
\end{split}
\end{align}

\paragraph{Full action}

Combining the two parts, the non-local and Chern-Simons terms, we get:
\begin{align}
\begin{split}
S^\star_{1} + S_2^\star = S^\star_{\tiny \mbox{SYM}} &=\int  \tr\left(B_{}\wedge_\star G_{_{ASD}} -\frac{g^2_{\tiny\mbox{YM}}}{2} B_{}\wedge_\star B_{}\right)\\
&\ \ \ -\frac{1}{4} \int \tr\left(D\Phi^{ij}\wedge_\star *D\Phi_{ij}\right) + \int d^2s d^2\bar s\ \tr\Big(\bar\Psi_i \star \sigma \wedge_\star *D\Psi^i \Big)\\
&\ \ \ + \int d^2\bar s \int\mathrm{d}^4x\ \tr\Big( \bar\Psi_i \star [\Phi^{ij}\stackrel{\star}{,}\bar\Psi_j] \Big) -g^2_{\tiny\mbox{YM}} \int d^2s\ \int \mathrm{d}^4x\ \tr\Big(\Psi^i\star[\Phi_{ij}\stackrel{\star}{,}\Psi^j]\Big)\\
&\ \ \ + \frac{g^2_{\tiny\mbox{YM}}}{4} \int \mathrm{d}^4x\ \tr\Big( [\Phi^{ij}\stackrel{\star}{,}\Phi^{kl}]\star[\Phi_{ij}\stackrel{\star}{,}\Phi_{kl}]\Big),
\end{split}
\end{align}
which after rescaling $\bar\Psi\to \sqrt{g_{\tiny\mbox{YM}}}\bar\Psi$ and $\Psi\to \Psi/\sqrt{g_{\tiny\mbox{YM}}}$, integrating out the auxiliary field $B$ yields the complete deformed $\mathcal{N}=4$ SYM action of the twisted noncommutative gauge theory as was presented in \cite{Meier:twist} up to adding a topological $\theta$-term, $\int\tr(G\wedge_\star G)$, that does not contribute perturbatively.

\section{Conclusions}
In this work we constructed gauge invariant actions on twisted twistor space with the aim of developing a proper mathematical framework for studying amplitude techniques in the setup of noncommutative gauge theories in four dimensions. We started by discussing the construction of a consistent twistor space twisted by a twist built from a Drinfel'd twist within the Poincar\'e algebra. In particular, we constructed a twist-deformed differential calculus on twistor space. We identified the twisted four-dimensional spacetime points in twisted twistor space as $\mathbb{CP}^1$ curves by imposing the same incidence relation as in the undeformed case. The restriction to Poincar\'e twists ensures that these curves remain intact. In \cite{Majid&Brain}, a comparable approach was developed in order to define a twisted version of Penrose transformations in the context of twisted twistor space. However, compared to the noncommutative twistor theories constructed from translational twists only, the coordinates on the residual $\mathbb{CP}^1$ do not commute anymore with spacetime coordinates. However, in the restricted case of twists in the Poincar\'e algebra, the residual $\mathbb{CP}^1$ coordinates still commute with each other, allowing for a standard integration over the fibre on the twisted twistor space. On these twisted twistor spaces we constructed noncommutative BF theory, which we showed to be equivalent to anti-self-dual Yang-Mills theory in four dimensions. Furthermore, by constructing a deformed version of the non-local term in the BF action, we were able to match with the full deformed Yang-Mills action in four dimensions discussed in \cite{Meier:twist,Meier:quad}. Beyond the pure bosonic Yang-Mills action we extended the construction to supertwistor space, yielding a noncommutative gauge-invariant holomorphic Chern-Simons action equivalent to the noncommutative version of maximally supersymmetric Yang-Mills theory (${\cal N}=4$ SYM), also considered in \cite{Meier:twist,Meier:quad}. For ensuring gauge invariance, we need to restrict to star products that are cyclic under integration. Our approach generalises the construction of gauge theories on constant noncommutative twistor space discussed in \cite{Takasaki:2000vs,Hamanaka:2005mq} to a broader class beyond the case corresponding to the Groenewold-Moyal case. Moreover, twists of only the internal R-symmetry as e.g. in the $\beta$ deformation \cite{Lunin}, which have been studied in \cite{Viana, Viana2} naturally fit our construction as an almost trivial example. We can also consider combined cases of twists built from Poincar\'e genrators and R-symmetry generators. This naturally includes dipole-like deformations, such as the light-like dipole case studied, for example, in \cite{Guica:2017mtd} and the angular dipole deformation in \cite{Meier:SpinChain}.

Furthermore, motivated by the noncommutative integrability in the Groenewold-Moyal twisted twistor space \cite{Takasaki:2000vs,Hamanaka:2005mq} and given the construction of the broader class of NC-deformed twistor theories in this work, it might be possible and valuable to extend the notion of NC integrability to this broader class of models. Furthermore, more recently a five-dimensional Groenewold-Moyal-like twistor theory has been shown to give rise to the most famous three-dimensional integrable system by reproducing the Lagrangian formulation of the KP equation in \cite{Bittleston:NCTwistor} which opens a new interesting direction of constructing higher dimensional integrability from twisted twistor spaces. Still in this direction, it is natural to further ask wether the twist-noncommutative deformations considered here induce corresponding deformations of the associated two-dimensional integrable gauge theories, along the lines developed in \cite{Ryan:Twistor,Ben:Diamond,Lewis:BFTheory,Thompson:Diamond} and references therein. In particular, an important open question is how such twists are reflected in the underlying integrable structures, for example through modified Lax connections. Furthermore, in \cite{Bittleston:2020hfv} and later in \cite{Ryan:Twistor,Ben:Diamond,Lewis:BFTheory,Thompson:Diamond} intrinsic connections between certain kinds of holomorphic 6d Chern-Simons theories, 4d Chern-Simons theories and four- and two-dimensional integrable field theories have been discovered. It would be very interesting to study similar connections in the setup of the kind of noncommutative twist deformations studied in this work.

Although this work focuses on finding a twistorial description of Poincar\'e twisted Yang-Mills theories, there is no direct obstruction in constructing gauge theories on twisted twistor space for twists of the complete $SL(4)$ symmetry of $\mathbb{CP}^3$. However, identification with four-dimensional spacetime models is more subtle for cases outside the Poincar\'e algebra as the fibre will not remain undeformed in these cases. Moreover, just recently, gauge invariant theories for twists including scale transformations have been constructed \cite{Borsato&Meier}. In this approach, the Hodge operator was modified. In particular, an auxiliary field was introduced in order to make manifest the scaling-dimension-changing behavior of the Hodge operator. It is interesting to extend our construction in this paper to such cases to give a better and more fundamental understanding of this auxiliary field in terms of twistor-space objects. Moreover, to this date a consistent noncommutative gauge-invariant Yang-Mills action including special conformal generators does not exist. A construction of the corresponding twistor theory could potentially overcome this problem and could be directly used to construct four-dimensional theories for twists of the full conformal algebra.

The most direct motivation for studying the YM theory via twistor theory is its natural formulation of MHV diagram rules for computing scattering amplitudes \cite{Witten,Mason:MHV,Adamo:MHV}. This provides a systematic and simpler way to calculate complex multi-particle processes while highlighting the geometric structure underlying gauge theory amplitudes. Having formulated a noncommutative version of twistor theory, the construction of deformed MHV rules might be a promising step towards calculating amplitudes in noncommutative Yang-Mills theories efficiently. Moreover, as physical observables, scattering amplitudes carry a representation of the symmetry algebra of the theory. As the algebra of the (super-)conformal symmetries gets twisted by the deformation, the scattering amplitudes of the noncommutative theory should carry a representation of the twisted symmetries. Together with the simplifying MHV techniques, this could shed light on how twisted symmetries are realized on physical observables such as scattering amplitudes and correlation functions.

\newpage

\section*{Acknowledgements}
We would like to thank Ben Hoare for valuable discussions and Lewis Cole and Stijn van Tongeren for valuable comments on the draft.
The work of TM was supported by the grant RYC2021-032371-I (funded by MCIN/AEI/10.13039/501100011033 and by the European Union “NextGenerationEU”/PRTR), the grant 2023-PG083 (with reference code ED431F 2023/19 funded by Xunta de Galicia), the grant PID2023-152148NB-I00 (funded by AEI-Spain), the María de Maeztu grant CEX2023-001318-M (funded by MICIU/AEI /10.13039/501100011033), the CIGUS Network of Research Centres, and the European Union.
The work of EV was supported in part by ICTP-SAIFR FAPESP grant 2019/21281-4 and by FAPESP grant 2022/00940-2.
\appendix

\section{Gauge invariance of the deformed non-local action on supertwistor space}\label{app:GaugeInvarianceNonLocal}
In order to show the gauge invariance of the non-local interaction, we will follow the same principles as in the undeformed case. Additionally, we will use the fact that the star product is cyclic under integration.

The non-local action completing the $\mathcal{N}=4$ SYM action is given by
\begin{equation}
S_2^\star[{\cal A}] = \sum_{n=2}^\infty\frac{1}{n}\left(\frac{1}{2\pi i}\right)^n\int_{(\mathbb{CP}^1)^n} \frac{\omega_1\cdots\omega_n}{\langle\lambda_1\lambda_2\rangle\cdots\langle\lambda_n\lambda_1\rangle}\tr\left({\cal A}_1\star\cdots\star{\cal A}_n\right),
\end{equation}
This action transforms under the gauge transformations $\delta_\epsilon{\cal A} = \bar\partial\epsilon + [\epsilon\stackrel{\star}{,}{\cal A}]$ as
\begin{align}
\begin{split}
\delta_\epsilon S_2^\star[{\cal A}] &=  \sum_{n=2}^\infty\frac{1}{n}\left(\frac{1}{2\pi i}\right)^n\int_{(\mathbb{CP}^1)^n} \frac{\omega_1\cdots\omega_n}{\langle\lambda_1\lambda_2\rangle\cdots\langle\lambda_n\lambda_1\rangle}\times\\
&\times\sum_{i_1}^n\Big(\tr\left(A_1\star\cdots\star[\epsilon_i\stackrel{\star}{,}{\cal A}_i]\star\cdots\star{\cal A}_n]\right)
+\tr\left(A_1\star\cdots\bar\partial\epsilon_i\star\cdots\star{\cal A}_n\right)\Big).
\end{split}
\end{align}
Although the twistor gauge parameter a priori depends on the fibre coordinate, $\epsilon_i(Z_i)$, any such dependence is pure twistor gauge, as a consequence of \eqref{eq:ResidualGauge}. Then, after gauge fixing along the fibres, $\epsilon_i$ will only depend on the spacetime coordinates $\epsilon_i(Z_i)=\epsilon(x)$. Then, after fixing the gauge along the fibres, the first term in the variation above reads:
\begin{align}
\sum_{n=2}^\infty\frac{1}{n}\left(\frac{1}{2\pi i}\right)^n\int_{(\mathbb{CP}^1)^n} \frac{\omega_1\cdots\omega_n}{\langle\lambda_1\lambda_2\rangle\cdots\langle\lambda_n\lambda_1\rangle}\Big( \tr\left([\epsilon\stackrel{\star}{,}{\cal A}_1\star{\cal A}_2\star\cdots\star{\cal A}_n]\right)\Big),
\end{align}
which is zero due to the cyclicity of the star product under integration and star-commutativity of the integral measures. For the second term, we note that only the zeroth components of the 1-forms contribute to this integral, so:
\begin{equation}
\delta_\epsilon S_2^\star[{\cal A}] =\sum_{n=2}^\infty\frac{1}{n}\left(\frac{1}{2\pi i}\right)^n\int_{(\mathbb{CP}^1)^n} \frac{\omega_1\cdots\omega_n}{\langle\lambda_1\lambda_2\rangle\cdots\langle\lambda_n\lambda_1\rangle}n\ \tr\Big(\bar\partial_0\epsilon_1\star({\cal A}_{2})_0\star \cdots \star({\cal A}_{n})_0\Big).
\end{equation}
Again, since $\epsilon_i$ can be taken to be independent of $\lambda_i$, $\bar\partial_0\epsilon_1=0$, and hence $\delta_\epsilon S_2^\star=0$.

\newpage

\providecommand{\href}[2]{#2}\begingroup\raggedright\endgroup

\end{document}